\documentclass[a4paper,fleqn,usenatbib]{mnras}

\usepackage[T1]{fontenc}
\usepackage{ae,aecompl}

\usepackage{newtxtext,newtxmath}
\usepackage{natbib}
\usepackage{hyperref}
\hypersetup{colorlinks,citecolor=Blue,linkcolor=Red,urlcolor=Blue}
\usepackage{amsmath}
\usepackage{empheq}
\usepackage[usenames,dvipsnames]{color}
\allowdisplaybreaks 

\newcommand{\appropto}{\mathrel{\vcenter{\offinterlineskip\halign{\hfil$##$\cr\propto\cr\noalign{\kern2pt}\sim\cr\noalign{\kern-2pt}}}}}

\newcommand{\Poincare}{{Poincar$\acute{\rm{e}}$}}
\newcommand{\Ham}{\mathcal{H}}
\newcommand{\G}{\mathcal{G}}
\newcommand{\LL}{\mathcal{L}}
\newcommand{\RR}{\mathcal{R}}
\newcommand{\II}{\mathcal{I}_\ell}

\newcommand{\ahat}{\hat{a}}
\newcommand{\abar}{\bar{a}}

\title[Secular Disks]{Schr\"{o}dinger Evolution of Self-Gravitating Disks}

\author[Konstantin Batygin]{
Konstantin Batygin\thanks{e-mail: kbatygin@gps.caltech.edu}
\\
Division of Geological and Planetary Sciences, California Institute of Technology, 1200 E. California Blvd., Pasadena, CA 91125, USA
}

\pubyear{2018}

\begin{document}
\label{firstpage}
\pagerange{\pageref{firstpage}--\pageref{lastpage}}

\maketitle

\begin{abstract}
An understanding of the long-term evolution of self-gravitating disks ranks among the classic outstanding problems of astrophysics. In this work, we show that the secular inclination dynamics of a geometrically thin quasi-Keplerian disk, with a surface density profile that scales as the inverse square-root of the orbital radius, are described by the time-dependent Schr\"{o}dinger equation. Within the context of this formalism, nodal bending waves correspond to the eigenmodes of a quasiparticle's wavefunction, confined in an infinite square well with boundaries given by the radial extent of the disk. We further show that external secular perturbations upon self-gravitating disks exhibit a mathematical similarity to quantum scattering theory. Employing this framework, we derive an analytic criterion for the gravitational rigidity of a nearly-Keplerian disk under external perturbations. Applications of the theory to circumstellar disks and Galactic nuclei are discussed. 
\end{abstract}


\section{Introduction}\label{sect1}

Astrophysical disks are among the most ubiquitous objects in the known universe. Generically understood to be a consequence of energy dissipation within gravitationally bound rotating systems, these structures span a staggering assortment of scales, ranging from galaxies to protoplanetary nebulae and circumplanetary rings \citep{GoldTrem1982}. Accordingly, characterizing the long-term evolution of self-gravitating disks constitutes one of the key challenges of dynamical astronomy.

Owing to their inherent diversity, astrophysical disks arise in nature with variable compositions, and can occupy rather distinct physical regimes. For example, active galactic nuclei and protoplanetary disks are predominantly composed of Hydrogen and Helium gas, and are representative of \textit{fluid} disks. As such, their evolution is governed by gravitational as well as (magneto-)hydrodynamic effects. Conversely, planetesimal/debris disks, as well as disks of stars that orbit supermassive black holes in the centers of galaxies (i.e. so-called \textit{particle} disks), are subject to essentially pure gravitational dynamics\footnote{Intriguingly, owing to their mildly collisional nature, Saturn's rings reside in between the ideal fluid and particle parameter regimes.} \citep{LatterReview2017}. 

Observed instances of astrophysical disks often encircle central objects that are much more massive than the disks themselves. The resulting dominance of the central body's gravitational potential leads to \textit{quasi-Keplerian} particle motion that resembles planetary orbits on short (orbital) timescales, but can exhibit non-trivial behavior over much longer periods of time, due to self-gravity \citep{Trem2001,ToumaTremaine2014}. Describing the long-term exchange of angular momentum within such systems is the primary aim of this paper. 

The specific goals of our analysis are essentially two-fold. Our first aim is to formulate a tangible theory for the secular inclination evolution of dynamically cold self-gravitating disks. Our second goal is to derive a dimensionless number, somewhat akin to Toomre's $Q$, that characterizes the propensity of disks to warp under external perturbations. In other words, we seek to obtain an analytical criterion for the \textit{gravitational rigidity} of astrophysical disks.

Quantifying the self-gravitational evolution of disks and their global response to external perturbations is essential to interpreting modern observations, as well as to the development of planet formation theory. In particular, the dynamical origins of the warped stellar disk at the center of our Galaxy have been attributed to a complex interplay between self-gravitational effects and torques from surrounding stellar clusters \citep{KocsisTremaine2011}. Likewise, large-scale warps and spiral morphology of young circumstellar disks\footnote{Well-known examples of extrasolar debris disks are found around Vega, Fomalhaut, $\beta$ Pictoris, and $\epsilon$ Eridani.} (see e.g. \citealt{2006AJ....131.3109G,2009ApJ...690.1522B} and the references therein) are routinely ascribed to interactions between the disks themselves and perturbing stellar or planetary companions \citep{Nesvold2016,Nesvold2017}. Even the seemingly unrelated question of the provenance of spin-orbit misalignments in exo-planetary systems may be deeply rooted in the process of gravitational torquing of protoplanetary nebulae by bound or passing stars \citep{2010MNRAS.401.1505B,SpaldingBatygin2014}.

Importantly, all of these phenomena arise on the so-called \textit{secular} timescale - one that greatly exceeds the characteristic orbital period, but is significantly shorter than the physical lifetime of the system (e.g. $\sim10^5\,$years). By operating in between the aforementioned temporal extrema, the dynamical mechanisms at play are poorly represented by either an explicit ($N$-body) description of orbital motion, or a diffusive model of angular momentum transfer within disks \citep{Pringle1981}. As a result, secular behavior of continuous self-gravitating systems remains imperfectly understood, and warrants the development of a simple theoretical model. 

At present, there exist three primary means of analyzing the dynamics of self-gravitating disks. The most straight-forward route is to employ direct $N$-body simulations. In spite of remarkable advances in computation that have transpired within the last decade \citep{2009NewA...14..630G,MooreQuillen2011}, this method remains too computationally expensive and specialized for most problems of interest. A second, more compact technique utilizes the collisionless Boltzmann equation, which, instead of granting $N$ individual trajectories as the solution, yields the evolution of the system's distribution function in phase-space \citep{BinneyTremaine1987,2017MNRAS.465.1856S}. A third approach relies on Gauss's averaging method of celestial mechanics, to replace the individual bodies with a series of massive wires that interact gravitationally among one-another on timescales much longer than the orbital period \citep{ToumaSoft2009,Batygin2012}. 

In this work, we focus exclusively on the third, mean-field model of self-gravitating disks. Specifically, working within the context of the Lagrange-Laplace secular perturbation theory, we demonstrate that the $N$-ring description of self-gravitating disks can be reduced to an evolution governed by Schr\"{o}dinger's equation. Correspondingly, this simplified framework allows for the computation of gravitational rigidity of astrophysical disks, and yields new insight into their long-term dynamical evolution. 

The paper is structured as follows. In section \ref{sect2}, we obtain Schr\"{o}dinger's equation as a continuum limit of Hamilton's equations. In section \ref{sect3}, we apply this calculation to the inclination dynamics of nearly-circular self-gravitating disks, and derive the corresponding eigenfunctions and eigenfrequencies. Importantly, we note that while the eigenfunctions of the disk can be approximately obtained by only considering nearest-neighbor interactions (section \ref{sect:soln}), the computation of the system's eigenfrequencies requires accounting for collective effects within the disk (section \ref{CEs}). In section \ref{sect4}, we extend this formalism to account for external perturbations, and deduce an analytic criterion for gravitational rigidity of perturbed systems. We summarize and discuss our results in section \ref{sect5}. 

\section{Schr\"{o}dinger's Equation From Hamilton's Equations}\label{sect2}

Prior to considering self-gravitating disks explicitly, let us deleniate a general framework that the calculations will follow. Recall that quantum and classical mechanics are routinely thought to be governed by Schr\"{o}dinger's and Hamilton's equations, respectively. Being macroscopic in nature, the latter is generally considered to be a limiting case of the former, where the action quantum $\hbar\rightarrow0$. This notion is strongly related to Ehrenfest's theorem, and is often referred to as the correspondence principle (see e.g. \citealt{1985mqm..book.....S}). In this section, we demonstrate the converse to also be true in a specific case. That is, for a particular Hamiltonian, Schr\"{o}dinger's equation naturally arises as a limiting case of Hamilton's equations.

\begin{figure}
\centering
\includegraphics[width=\columnwidth]{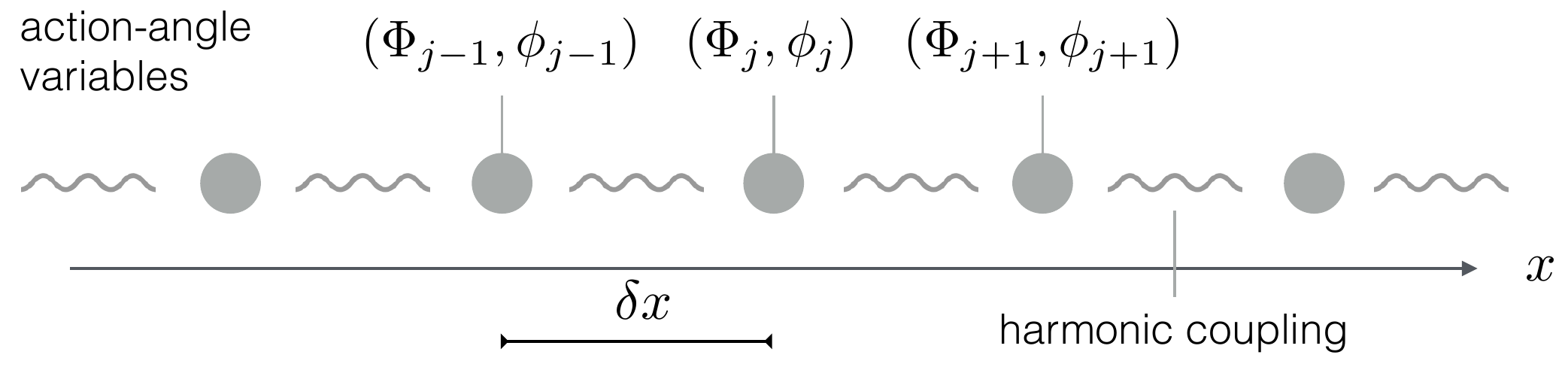}
\caption{An infinite chain of harmonically coupled oscillators. The positions of the objects are fixed on the $x-$axis, such that the distance between neighboring sites is $\delta x$. The dynamical state of each oscillator is described by a pair of action-angle coordinates, $(\Phi,\phi)$.}
\label{nodesfig}
\end{figure}

Consider an infinite sequence of coupled objects lying on the $x$-axis, whose individual dynamical state is described by a pair of action-angle variables $... \,(\Phi_{j-1},\phi_{j-1})$, $(\Phi_{j},\phi_{j})$, $(\Phi_{j+1},\phi_{j+1})\, ...\,$. Let the spacing between the objects be equidistant, and denote it $\delta x$ (Figure \ref{nodesfig}). Finally, let the purely classical Hamiltonian describing the evolution of object $j$ be
\begin{align}
\Ham &= 2\,c_1\,\Phi_j+2\,c_2\,\sqrt{\Phi_j\,\Phi_{j+1}} \cos(\phi_j-\phi_{j+1}) \nonumber \\
&+2\,c_2\,\sqrt{\Phi_j\,\Phi_{j-1}} \cos(\phi_j-\phi_{j-1}),
\label{H1}
\end{align}
where $c_1$ and $c_2$ are quantities that can depend on $x$ and $t$. Note that in the above expression, only interactions between the nearest neighbors are considered. 

Next, we define canonical cartesian analogs $(\xi,\zeta)$ to the action-angle variables $(\Phi,\phi)$:
\begin{align}
&\xi=\sqrt{2\,\Phi}\cos(\phi) &\zeta=\sqrt{2\,\Phi}\sin(\phi),
\end{align}
and organize them into a single complex coordinate:
\begin{align}
\Psi = \frac{\xi+\imath\,\zeta}{\sqrt{2}} = \sqrt{\Phi} \exp(\imath\,\phi),
\label{complexPsi}
\end{align}
where $\imath=\sqrt{-1}$. The Hamiltonian then reads:
\begin{align}
\Ham &= 2\,c_1\,\Psi_j \,\Psi^*_j+2\,c_2\,\big(\Psi_j \, \Psi^{*}_{j+1} + \Psi^{*}_{j}  \Psi_{j+1}+ \Psi_j \, \Psi^{*}_{j-1} \nonumber \\
&+ \Psi^{*}_{j}  \Psi_{j-1}\big),
\label{H20}
\end{align}
where $\Psi^{*}$ is a complex conjugate to $\Psi$.

In terms of complex variables, Hamilton's equations take on a rather succinct form \citep{Strocchi1966}:
\begin{align}
\frac{d\Psi_j}{dt} = \imath\,\frac{\partial \Ham}{\partial \Psi_j^{*}} = \imath \, \bigg(2\,c_1\,\Psi_j +c_2 \left(\Psi_{j+1} + \Psi_{j-1} \right) \bigg).
\label{EOM}
\end{align}
If we envision $\Psi_{j-1},\Psi_{j},\Psi_{j+1},...$ to be a discrete representation of a continuous, complex field $\Psi$, we may write down the central difference approximation for its second derivative as follows:
\begin{align}
\frac{\partial^2\,\Psi_{j}}{\partial\,x^2} \simeq\frac{\Psi_{j+1} - 2\,\Psi_j  + \Psi_{j-1}}{\left(\delta x\right)^2}.
\label{second}
\end{align}
We note that a similar approximation is routinely made in the elastic continuum representation of harmonic crystals \citep{Animalu1977}, and corresponds to the limit where the wave numbers of interest greatly exceed the spacing between atomic sites (i.e. $2\pi/k\gg\delta x$).

Plugging in equation (\ref{second}) into equation (\ref{EOM}) and multiplying both sides by $-\hbar/\imath$, we obtain:
\begin{align}
\imath\,\hbar \, \frac{\partial \Psi_j}{\partial t} = -\hbar\,\bigg(2(c_1+c_2)\Psi_{j} + c_2 \left(\delta x \right)^2 \frac{\partial^2\,\Psi_j}{\partial\,x^2} \bigg).
\label{almost}
\end{align}
Note that here, we have taken advantage of the fact that the dynamical objects are fixed in their $x$-coordinate to set $d\Psi/dt=\partial \Psi /\partial t$. Adopting the following expressions for the interaction constants:
\begin{align}
&c_1 = -\frac{V(x,t)}{2}- \frac{\hbar}{2\,\mu \left(\delta x \right)^2}   &c_2 = \frac{\hbar}{2\,\mu \left(\delta x \right)^2},
\label{constants}
\end{align}
and taking the limit as $\delta x\rightarrow 0$ (which renders the relationship in equation (\ref{second}) exact), we obtain the time-dependent Schr\"{o}dinger's equation:
\begin{align}
\imath\,\hbar \,\frac{\partial\, \Psi(x,t)}{\partial t} = \left[ -\frac{\hbar^2}{2\,\mu} \frac{\partial^2}{\partial\,x^2}  + V(x,t) \right] \Psi(x,t).
\label{done}
\end{align}

\begin{figure*}
\centering
\includegraphics[width=\textwidth]{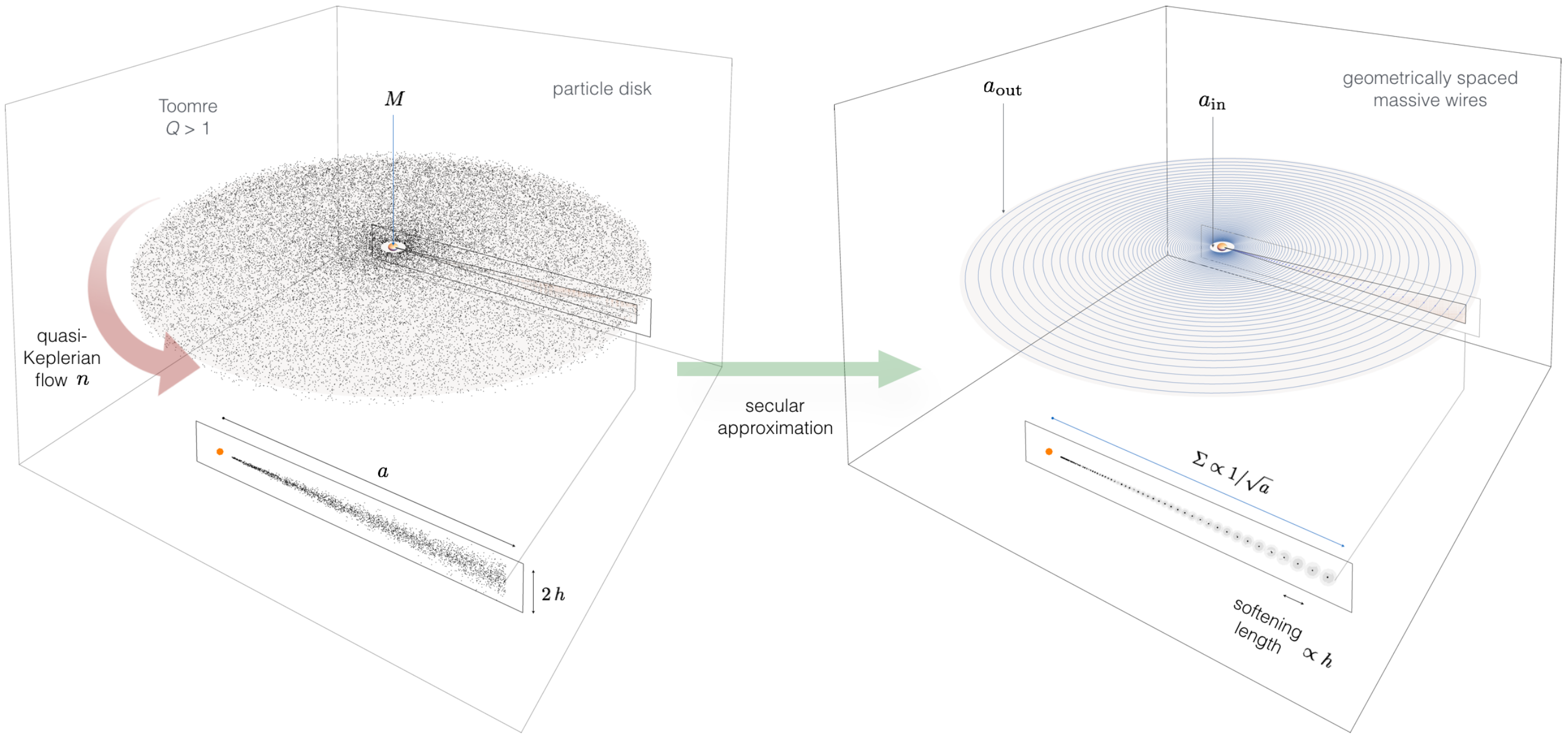}
\caption{Geometric setup of our model. A quasi-Keplerian disk of $N\gg1$ particles is modeled as a sequence of geometrically spaced massive wires. The gravitational potential of each wire is softened by the aspect ratio of the disk, accounting for the inherent velocity dispersion of the constituent particles. A surface density profile that scales inversely with the square root of the semi-major axis is assumed.}
\label{setupfig}
\end{figure*}

This derivation is trivially generalizable to three spatial dimensions. Moreover, a simple modification of the same procedure naturally leads to the nonlinear variant of Schr\"{o}dinger's equation. That is, addition of a nonlinear action term to equation (\ref{H1}), such that 
\begin{align}
\Ham' = \Ham + \frac{\kappa}{2}\,\Phi_j^2,=\Ham+\frac{\kappa}{2}\,(\Psi_j\,\Psi_j^*)^2
\label{H2}
\end{align}
yields 
\begin{align}
\imath\,\hbar \frac{\partial\, \Psi(x,t)}{\partial t} &= \left[ -\frac{\hbar^2}{2\,\mu} \frac{\partial^2 }{\partial\,x^2}  + V(x,t) \right] \Psi(x,t) \nonumber \\
&+ \kappa\,|\Psi(x,t)|^2\,\Psi(x,t).
\label{done2}
\end{align}
It is worth noting that a Hamiltonian of the form (\ref{H2}) is often referred to as the ``second fundamental model of resonance,'' and possesses a phase-space topology similar to that of a harmonic oscillator or a pendulum, depending on the assumed values of the constants \citep{1983CeMec}.

This work is by no means the first to derive Schr\"{o}dinger's equation from purely classical principles. For example, \citet{Nelson1966} showed that Schr\"{o}dinger's equation can be obtained as a description of a stochastic system, where particles are subjected to Brownian motion with a diffusion coefficient $\hbar/2m$, in addition to Newton's laws of motion. Extension of this formalism to an effective field theory has been considered by \citet{Guerra1981}, and has led to the so-called stochastic interpretation of quantum mechanics. 

In contrast with these works, here we did not draw on stochastic fluctuations to augment classical mechanics. At the same time, we emphasize that even though the above derivation is self-consistent, it is not general, and only applies to a particular Hamiltonian that describes a continuous chain of forced harmonic oscillators with specific interaction coefficients. Incidentally, the governing equations of Largrange-Laplace secular theory of celestial mechanics can be cast into this form. 

\section{Secular Evolution of Self-Gravitating Disks}\label{sect3}

Having outlined the general course of action in the previous section, we now proceed to show how the Schr\"{o}dinger equation can be used to understand the long-term evolution of self-gravitating disks. However, prior to computing the dynamical evolution itself, it is necessary to define a number of basic properties of our model. 

\subsection{Disk Profile} \label{sectdiskprofile}

Consider a disk of material, comprised of a large number of point masses in orbit around a single central body of mass $M$. Envision that the gravitational potential of the central body dominates, such that the trajectories of the individual bodies follow Kepler's third law: 
\begin{align}
n = \sqrt{\frac{\G M}{a^3}},
\label{Kepsecond}
\end{align}
where $n$ is the mean motion, and $a$ is the semi-major axis. Further, assume that the orbital eccentricities as well as mutual inclinations of neighboring orbits are not exceedingly large, meaning that the inherent velocity dispersion of a population of objects occupying the same semi-major axis range is modest compared to the Keplerian velocity. Correspondingly, we define the aspect ratio of the disk, $\beta$, as an intrinsic small parameter of the problem, and take it to be constant:
\begin{align}
\beta \equiv \frac{h}{a} = \rm{const.} \ll 1,
\label{beta}
\end{align}
where $h$ is the characteristic scale-height. Crucially, the value of $\beta$ sets the gravitational softening length of the disk\footnote{Intuitively, softening the gravitational potential of a point-mass by a length $h$ can be thought of as spreading the mass of the object over a Plummer sphere of radius $h$.} \citep{Adams1989,2002MNRAS.333..583T}.

Let us now replace this collection of secondary bodies with a series of $N$ massive ``streams,'' thus turning the aggregate of orbiting particles into a disk comprised of nested elliptical wires, with normalized thickness $\beta$ (Figure \ref{setupfig}). Importantly, we assume that the wires do not cross and that the epicyclic motion of constituent particles (i.e. distribution of eccentricities at a given semi-major axis) is fully encapsulated by softening parameter \citep{BinneyTremaine1987}. The spacing of the wires is taken to be geometric, such that the semi-major axis ratio of neighboring orbits, $\alpha$, is constant, and roughly corresponds to a single softening length:
\begin{align}
\alpha \equiv \frac{a_{j-1}}{a_j} = \frac{a_{j}}{a_{j+1}} = \frac{1}{1+\beta} \approx 1.
\label{alpha}
\end{align}
In light of this definition, it is useful to scale the semi-major axis by the disk's inner truncation radius, $a_{\rm{in}}$, and introduce a dimensionless logarithmic radial coordinate
\begin{align}
&\rho \equiv \log\bigg( \frac{a}{a_{\rm{in}}} \bigg) &\mathcal{L} \equiv \log\bigg( \frac{a_{\rm{out}}}{a_{\rm{in}}} \bigg),
\label{rho}
\end{align}
where $a_{\rm{out}}$ denotes the outer boundary of the disk. In terms of this quantity, the wires are spaced equidistantly, and the disk boundaries are given by $\rho\in[0,\LL]$. To this end, we note that any realistic system will have $\LL$ on the order of a few, rarely exceeding 10. Coincidentally, for a typical protoplanetary neblua (e.g. $a_{\rm{in}} \sim 0.05\,$AU and $a_{\rm{out}} \sim 50\,$AU), $\LL \sim 2\,\pi$.

Qualitatively speaking, the process of ``smearing out'' individual bodies into massive rings, whose line densities are inversely proportional to orbital velocity, is equivalent to canonically averaging over the rapidly-varying orbital angles (i.e. mean longitudes). This procedure yields orbital semi-major axes, $a$, that are frozen in time, because the action conjugate to mean longitude is proportional to $\sqrt{a}$ \citep{morbybook}. In turn, this conservation implies that the 2-body (Keplerian) energy of each object is preserved, and the wires only exchange angular momentum. Physically, this process leads to slow precession of the apsidal and nodal lines of the orbits as well as variations in eccentricities and inclinations. 

A conventional parameterization of density variations in astrophysical disks assumes that the surface density, $\Sigma$, scales as some negative power of the orbital radius \citep{Armitage2010}. Here, we follow this standard prescription and adopt an inverse square root surface density profile of the disk:
\begin{align}
\Sigma = \Sigma_0 \, \bigg(\frac{a_0}{a}\bigg)^{1/2},
\label{sigmaprofile}
\end{align}
where $\Sigma_0$ is the surface density at a reference semi-major axis, $a_0$. We note that while this profile is somewhat shallower than, say, a classical \citet{1963MNRAS.126..553M} disk which takes $\Sigma \propto 1/a$, it is routinely employed in numerical studies of protoplanetary disks (e.g. \citealt{2011A&A...536A..77B,2017A&A...606A.146L}). Strictly speaking, this approximation is not necessary. However, here we choose to adopt expression (\ref{sigmaprofile}) for illustrative purposes, as it will simplify some of the calculations down the line. 

With the above definitions in place, we obtain the following expression for the mass of an individual wire (of index $j$):
\begin{align}
m_j &= \oint \int_{a(1-\beta/2)}^{a(1+\beta/2)} \Sigma\,a\,da\,d\varphi \approx 2\,\pi\,\beta\,\Sigma_0\sqrt{a_0\,a_j^3}+\mathcal{O}(\beta^3),
\label{mj}
\end{align}
where $\varphi$ denotes the azimuthal coordinate. Correspondingly, assuming that $a_{\rm{in}}\ll a_{\rm{out}}$, the total disk mass is given by:  
\begin{align}
m_{\rm{disk}} &= \oint \int_{a_{\rm{in}}}^{a_{\rm{out}}} \Sigma\,a\,da\,d\varphi \approx \frac{4\,\pi}{3}\,\Sigma_0\,\sqrt{a_0\,a_{\rm{out}}^3}.
\label{mdisk}
\end{align}

The approximations employed above are only sensible if the dynamical evolution that ensues on the orbital timescale does not deviate significantly from purely Keplerian motion. Quantitatively, this statement corresponds to the requirement of the gravitational stability of the system \citep{Safronov1960,Toomre1964}:
\begin{align}
Q = \frac{h\,n^2}{\pi\,\G\,\Sigma} \gtrsim 1.
\label{Qtoomre}
\end{align}
In practical terms, this restriction translates to an upper-limit on $m_{\rm{disk}}$:
\begin{align}
m_{\rm{disk}} \lesssim \beta \, \frac{M}{2},
\label{mdiskbeta}
\end{align}
and yields a connection between the inherent velocity dispersion of constituent matter and surface density of the disk. With the preliminary specifications of the model now in place, we now continue on to compute the dynamical evolution.

\subsection{Inclination Dynamics}

Lagrange-Laplace secular theory constitutes one of the earliest, and best-known results of perturbation theory in celestial mechanics. Within the framework of this model, the phase-averaged gravitational potential of the interacting bodies (i.e. the negative \textit{disturbing function}) is expanded as a Fourier series in the orbital angles and as a power-series in eccentricities and inclinations \citep{EllisMurray2000}. Truncating the expansion at second order in both quantities yields secular vibrations of inclinations that are decoupled from the oscillations of eccentricities\footnote{Coupling terms arise at fourth order in the perturbation series, and are responsible for secular chaos within the solar system \citep{1994A&A...287L...9L,2015ApJ...799..120B}.}. Taking advantage of this disconnect between the degrees of freedom, here we treat the inclination dynamics, simply assuming that the eccentricities are small.

\subsubsection{Governing Equation}
Consider the inclination dynamics of a wire (labeled by index $j\neq1,N$) embedded within a radially extended disk. As a first approximation, let us restrict the range of interactions of wire $j$ to its nearest neighbors, $j+1$ and $j-1$. This approximation has the obvious shortcoming of reduced coupling within the disk, and we will revisit (and fix) this limitation later in the manuscript. The relevant (scaled) disturbing function\footnote{The unscaled disturbing function is a measure of energy, like the Hamiltonian. Reduced by a characteristic angular momentum $m\sqrt{\G\,M\,a}$, the scaled disturbing function becomes a measure of frequency (and has units of inverse time). For the remainder of the paper, we will refer to the scaled disturbing function as simply the disturbing function.} that governs the exchange of angular momentum is \citep{MD99}:
\begin{align}
\RR_j &= \frac{1}{2}B_{jj} \, i_j^2 + B_{jj-1} \, i_j\,i_{j-1} \cos(\Omega_j - \Omega_{j-1}) \nonumber \\
&+ B_{jj+1} \, i_j\,i_{j+1} \cos(\Omega_j - \Omega_{j+1}),
\label{Ri}
\end{align}
where $i$ is orbital inclination, $\Omega$ is the longitude of ascending node, and $B$'s are interaction coefficients that depend only on the semi-major axis ratios and masses. 

Although Keplerian elements $i$ and $\Omega$ do not constitute a set of canonically conjugated variables, the quantities 
\begin{align}
&p=i\,\sin(\Omega) &q=i\,\cos(\Omega)
\label{pq}
\end{align}
do, where $p$ is interpreted as the coordinate, and $q$ is the momentum (with $\RR_j$ acting as the Hamiltonian). Note that physically, $p$ and $q$ represent a measure of the angular momentum \textit{deficit} of wire $j$ in the $\hat{z}-$direction. Collecting these variables into a single complex coordinate akin to equation (\ref{complexPsi}):
\begin{align}
\eta=\frac{q+\imath\,p}{\sqrt{2}}=\frac{i}{\sqrt{2}}\exp\big(\imath \, \Omega \big),
\label{eta}
\end{align}
we rewrite equation (\ref{Ri}) as follows:
\begin{align}
\RR_j &= B_{jj} \, \eta_j \,\eta^*_j + B_{jj-1} \, \big(\eta_j \,\eta^*_{j-1}+\eta_{j-1} \,\eta^*_{j} \big) \nonumber \\
&+ B_{jj+1} \, \big(\eta_j \,\eta^*_{j+1}+\eta_{j+1} \,\eta^*_{j} \big).
\label{Rieta}
\end{align}

This expression is clearly reminiscent of Hamiltonian (\ref{H20}). However, in order to draw further analogy between equations (\ref{Rieta}) and (\ref{H20}), we must evaluate the relationship between the coefficients $B_{jj}$, $B_{jj-1}$, and $B_{jj+1}$. Under the assumption that $m_j\ll M$, these quantities are expressed as follows \citep{MD99}:
\begin{align}
&B_{jj-1}=\frac{n_j}{4}\frac{m_{j-1}}{M}\,\alpha\,\tilde{b}_{3/2}^{(1)}\{ \alpha \} \nonumber \\
&B_{jj+1}=\frac{n_j}{4}\frac{m_{j+1}}{M}\,\alpha^2\,\tilde{b}_{3/2}^{(1)} \{ \alpha \} \nonumber \\
&B_{jj}=-\big( B_{jj-1}+B_{jj+1} \big),
\label{Bcoeffs}
\end{align}
where $\tilde{b}_{3/2}^{(1)}\{ \alpha \}$ is the Laplace coefficient of the first kind. Recalling from equation (\ref{alpha}) that $\alpha \approx 1-\beta$, expressions (\ref{Bcoeffs}) imply $B_{jj-1} \approx B_{jj+1}$. In particular, upon defining the quantity
\begin{align}
\mathcal{B}&=\frac{n_j}{4}\frac{m_{j}}{M}\,\frac{\alpha}{2}\,\tilde{b}_{3/2}^{(1)} \{ \alpha \} \bigg[ \frac{m_{j-1	}}{m_j}+\frac{m_{j+1}}{m_j}\frac{1}{1+\beta} \bigg] \nonumber \\
&=\frac{n_j}{4}\frac{m_{j}}{M}\,\alpha\,\tilde{b}_{3/2}^{(1)} \{ \alpha \} \bigg[ \frac{2+2\,\beta+\beta^2}{2\,(1+\beta)^{3/2}} \bigg],
\label{Bcommon}
\end{align}
we have:
\begin{align}
& \lvert B_{jj-1} - \mathcal{B} \rvert = | B_{jj+1} - \mathcal{B} | \approx \frac{\beta}{4} + \mathcal{O}(\beta^2) \ll 1 \nonumber \\
&B_{jj} = - 2\,\mathcal{B}.
\label{Bdiff}
\end{align}

At this point, the advantage of choosing the specific form of the surface density profile (\ref{sigmaprofile}) becomes evident. Since $m_j\propto\sqrt{a^3}$ and $n_j\propto1/\sqrt{a^3}$, the two dependencies cancel, rendering $\mathcal{B}$ constant throughout the disk. Thus, to an excellent approximation,
\begin{align}
\frac{d\eta_j}{dt} = \imath \frac{\partial\RR_j}{\partial \eta_j^*}\approx \imath\, \mathcal{B}\, \big(\eta_{j-1}-2\eta_j+\eta_{j+1} \big).
\label{EOMinc1}
\end{align}
We note that under a different choice for the radial dependence of the surface density profile, the quantity $\mathcal{B}$ would become an explicit function of the semi-major axis. 

Because the disk annuli are spaced geometrically in $a$, they are equidistant in $\rho$ (equation \ref{rho}), with
\begin{align}
\delta\rho=\rho_{j+1}-\rho_{j}=\rho_{j}-\rho_{j-1}=\log(1+\beta).
\label{deltaroh}
\end{align}
Hence, employing the finite-difference approximation (equation \ref{second}) and recalling that the semi-major axes of the wires are secularly invariant, the continuum limit of equation (\ref{EOMinc1}) takes the form:
\begin{align}
\frac{\partial\eta}{\partial t} = \imath\, \mathcal{B}\,\big(\log(1+\beta) \big)^2\,\frac{\partial^2\eta}{\partial\rho^2}.
\label{EOMinccont}
\end{align}

Due to the proximity of the neighboring wires to each-other, it is sensible to evaluate equation (\ref{Bcommon}) in the limit where $\alpha\rightarrow1$. However, Laplace coefficients are notoriously singular at $\alpha=1$ (this is simply a re-statement of the fact that the gravitational potential becomes infinite at null separations). This mathematical inconvenience is easily circumvented by accounting for the inherent velocity dispersion of particles within the disk and softening the Laplace coefficient by the disk aspect ratio \citep{2002MNRAS.333..583T,2003ApJ...595..531H}:  
\begin{align}
\tilde{b}_{\ell}^{(j)} \{ \alpha \}=\frac{2}{\pi}\int_{0}^{\pi}\frac{\cos(j\,\psi)}{\big(1-2\,\alpha\cos(\psi)+\alpha^2+\beta^2 \big)^{\ell}}\, d\psi.
\label{Laplacecoeff}
\end{align}
These softened Laplace coefficients are well-behaved in the $\alpha\rightarrow1$ limit, and the quantity $\tilde{b}_{3/2}^{(1)}$ can now be expressed in terms of standard elliptic integrals. For the time being, it suffices to evaluate this function at $\alpha=1/(1+\beta)$ to obtain:
\begin{align}
\alpha\,\tilde{b}_{3/2}^{(1)} \{ \alpha \} \approx \frac{1}{\pi\,\beta^2} + \mathcal{O}(\beta^{-1}).
\label{feexp}
\end{align}

\begin{figure}
\centering
\includegraphics[width=\columnwidth]{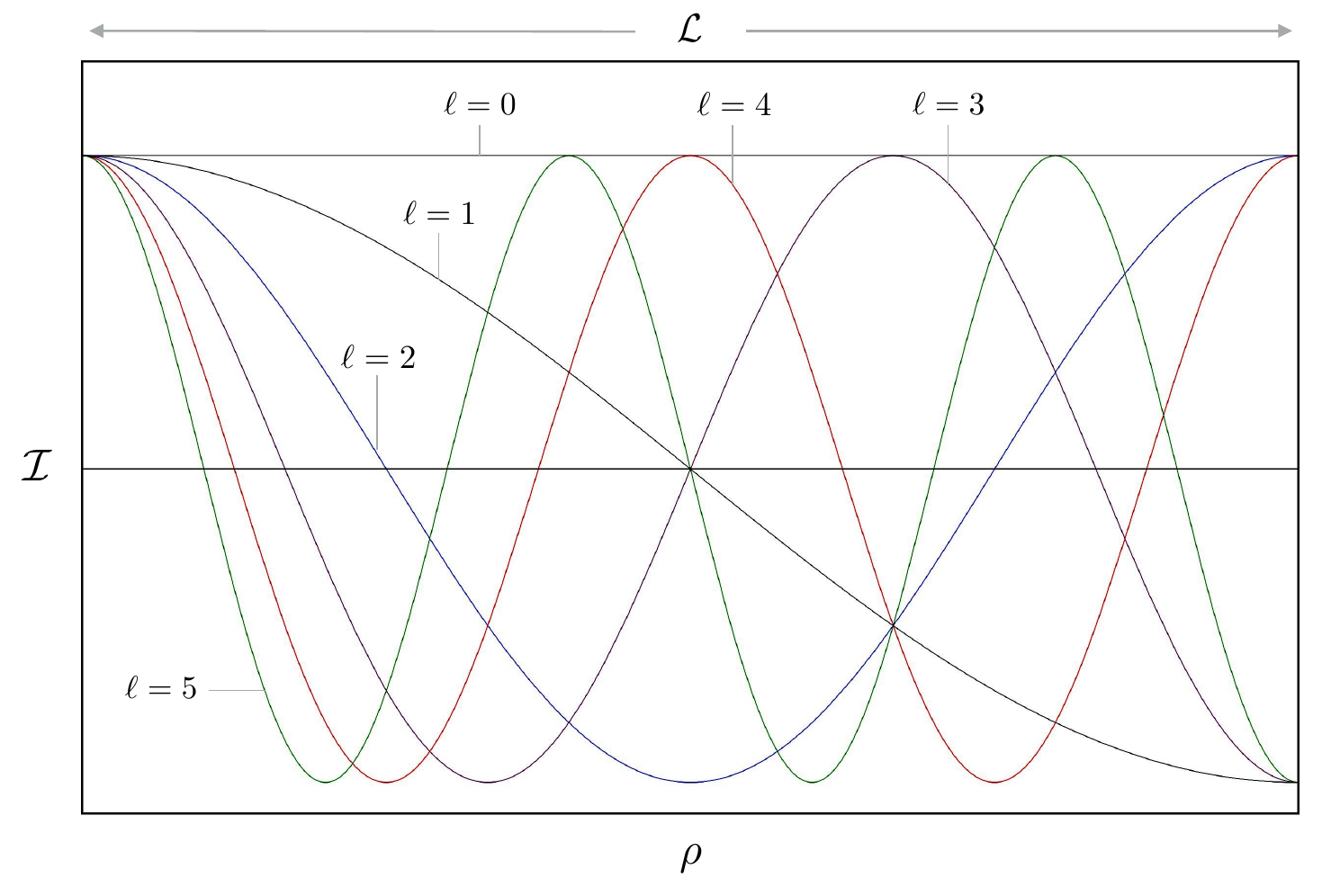}
\caption{Normal modes of a razor-thin ($\beta\rightarrow0$) disk. Six low-frequency modes are shown, and are labeled by the appropriate index $\ell=0,1,...5$. Stationary modes of a geometrically thin disk with $\beta=0.01$ appear nearly indistinguishable from those depicted in this figure.}
\label{modesincfig}
\end{figure}

Substituting this result back into equation (\ref{Bcommon}),  multiplying both sides by $\imath\,\omega_i$, while expanding to leading order in $\beta$, and noting that $\log(1+\beta)\approx\beta$, we obtain the potential-free Schr\"{o}dinger's equation of inclination dynamics within the disk
\begin{empheq}[box=\fbox]{align}
& \imath\, \omega_i \, \frac{\partial\eta}{\partial t}  = - \omega_i^2 \,\frac{\partial^2\eta}{\partial\rho^2},
\label{Schrodinc}
\end{empheq}
where
\begin{align}
& \omega_i = \frac{n}{4\,\pi}\frac{m}{M} = \frac{\beta\,\Sigma_0 \sqrt{\G M a_0}}{2 M}.
\label{omegai}
\end{align}
Thus, the mathematical description of secular angular momentum exchange within self-gravitating disks parallels that of a quantum particle confined to an infinite square potential well, with boundaries that extend from $\rho=0$ to $\rho=\mathcal{L}$. Note further that in this formalism, the fundamental frequency $\omega_i$ - a quantity related to the angular momentum budget of the disk - takes the place of $\hbar$\footnote{We remind the reader that assuming a functional form for the surface density profile other than that given by equation (\ref{sigmaprofile}) would render $\omega_i$ a function of $a$.}.

From intuitive grounds, one may speculate that because we have limited ourselves to only considering near-neighbor interactions in the treatment outlined above, $\omega_i$ must substantially under-estimate the true wave propagation frequency within the disk. This is indeed the case, and the limitations arising from this approximation will be addressed quantitatively in section \ref{CEs}. In the meantime, however, it is advantageous to temporarily ignore this problem and examine the character of the normal modes dictated by the above equation.

\begin{figure*}
\centering
\includegraphics[width=\textwidth]{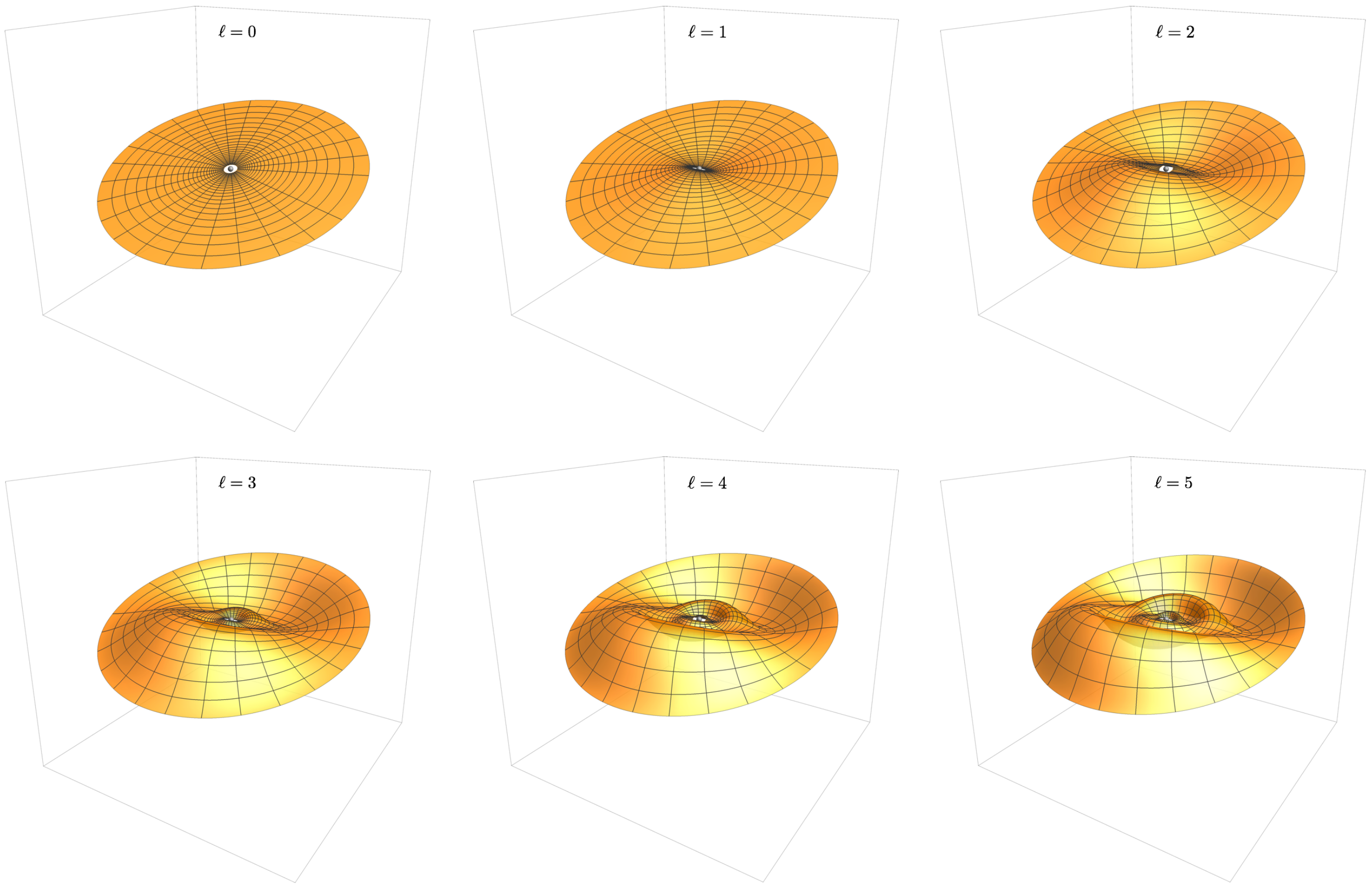}
\caption{Six low-frequency inclination normal modes of a razor-thin disk, rendered in physical space. Note that owing to the definition of the logarithmic coordinate $\rho$, large-scale variations associated with higher mode indexes are concentrated towards small orbital radii. Here, mode amplitudes of $c_\ell=\pi/10$ are assumed throughout.}
\label{incmodesphysfig}
\end{figure*}

\subsubsection{Boundary Conditions} \label{BCS}
The specific character of the solution to equation (\ref{Schrodinc}) is determined by the imposed boundary conditions. In the well-known problem of a quantum-mechanical infinite square well, it is appropriate to employ the Dirichlet boundary conditions, $\eta=0$ at $\rho=0$ and $\rho=\mathcal{L}$, since the wavefunction must vanish at the boundaries. On the contrary, for the problem at hand, there is no reason to require the orbital inclination to approach a particular value at the margins of the disk, and instead the boundary conditions must be deduced from the behavior of the discrete system at the disk's inner and outer edges.

Let us consider the outer edge first. The disturbing function for the wire indexed by $j=N$ is
\begin{align}
\RR_N &= - \mathcal{B} \, \eta_N \,\eta^*_N + \mathcal{B} \, \big(\eta_N \,\eta^*_{N-1}+\eta_{N-1} \,\eta^*_{N} \big),
\label{RNeta}
\end{align}
where we have recycled the same approximations from the preceding sub-section. The resulting equation of motion has the form:
\begin{align}
\frac{d\eta_N}{dt} &= \imath\frac{\partial \RR_N}{\partial \eta_N^*} = - \imath \, \mathcal{B} \big( \eta_N - \eta_{N-1} \big).
\label{RNetaEOM}
\end{align}
The RHS of this expression is a finite difference approximation for the first derivative of $\eta$ with respect to $\rho$. Correspondingly, recalling that $\mathcal{B} \, (\delta\rho)\approx\omega_i/\beta$, we obtain the boundary condition at $\rho=\mathcal{L}$:
\begin{align}
\imath\frac{\partial\eta}{\partial \rho} &=  - \frac{\beta}{\omega_i} \frac{\partial\eta}{\partial t} .
\label{BCrhoL}
\end{align}

An identical procedure can be carried out for the inner boundary by change of indexes. Specifically, the appropriate disturbing function is
\begin{align}
\RR_1 &= - \mathcal{B} \, \eta_1 \,\eta^*_1 + \mathcal{B} \, \big(\eta_1 \,\eta^*_{2}+\eta_{2} \,\eta^*_{1} \big).
\label{RNetain}
\end{align}
However, because the order of the wires is now reversed, the derivative appears with the opposite sign. The boundary condition at $\rho =0$ is thus
\begin{align}
\imath\frac{\partial\eta}{\partial \rho} &=  \frac{\beta}{\omega_i} \frac{\partial\eta}{\partial t}.
\label{BCrho0}
\end{align}
With these specifications in place, we can now proceed to write down the solution to the governing equation (\ref{Schrodinc}).

\subsubsection{Solution} \label{sect:soln}
Let us begin by noting that the potential-free Schr\"{o}dinger equation (\ref{Schrodinc}) itself is a diffusion equation in imaginary time. Physically, this means that the ``diffusion'' process must conserve the phase-space volume of the underlying distribution function (i.e. the time evolution must be unitary; \citealt{Stonebook}). This is perfectly sensible since in section (\ref{sect2}) we derived Schr\"{o}dinger's equation from Hamilton's equations, which are themselves rooted in Liouville's theorem. Accordingly, rather than describing decay as its solution, equation (\ref{Schrodinc}) must instead be satisfied by standing waves - i.e., normal modes of a disk.

Let $\ell$ denote the index of a mode (characterized by frequency $\omega_\ell$). By separation of variables
\begin{align}
\eta_{\ell}=c_\ell\,\exp(-\imath\,\omega_\ell\, t)\,\II,
\label{etasol}
\end{align}
where $c_\ell$ is a constant, we obtain the familiar harmonic oscillator equation
\begin{align}
\omega_\ell\, \II+\omega_i\frac{\partial^2\II}{\partial\rho^2}=0,
\label{Iinceqn}
\end{align}
subject to the boundary conditions
\begin{align}
&\frac{\partial \II}{\partial \rho} = -\beta\bigg(\frac{\omega_\ell}{\omega_i} \bigg) \II \ \bigg|_{\rho=0} &\frac{\partial \II}{\partial \rho} = \beta\bigg(\frac{\omega_\ell}{\omega_i} \bigg) \II \ \bigg|_{\rho=\mathcal{L}}.
\label{Ibounds}
\end{align}
Note that for long-wavelength modes, where $\omega_\ell$ does not exceed the fundamental frequency $\omega_i$ by a large margin, expressions (\ref{Ibounds}) are well approximated by the rudimentary Neumann boundary conditions $\partial\II/\partial \rho = 0$ at $\rho=0,\mathcal{L}$, characteristic of a string with free ends.

The solution to equation (\ref{Iinceqn}) is 
\begin{align}
\II = \cos\bigg( \sqrt{\frac{\omega_\ell}{\omega_i}} \rho \bigg) - \beta \sqrt{\frac{\omega_\ell}{\omega_i}} \sin \bigg( \sqrt{\frac{\omega_\ell}{\omega_i}} \rho \bigg),
\end{align}
with the quantization of $\omega_\ell/\omega_i$ stemming from the condition
\begin{align}
\bigg[\frac{\beta \,\omega_\ell}{2\,\omega_i} -\frac{1}{2\, \beta} \bigg]\sin \bigg( \sqrt{\frac{\omega_\ell}{\omega_i}} \mathcal{L} \bigg)= \sqrt{\frac{\omega_\ell}{\omega_i}} \cos\bigg( \sqrt{\frac{\omega_\ell}{\omega_i}} \mathcal{L} \bigg).
\label{quantizenoGR}
\end{align}
Although this expression does not admit a simple solution, in the $\beta\rightarrow0$ limit of a razor-thin disk, it simplifies to $\sin\left(\sqrt{\omega_\ell/\omega_i}\,\mathcal{L}\right)=0$, which is trivially solvable. Adopting the razor-thin limit as a leading-order approximation, it is straightforward to derive a correction to the quantization condition:
\begin{align}
\frac{\omega_\ell}{\omega_i}=\bigg(\frac{\ell\,\pi}{\mathcal{L}} \bigg)^2\bigg(1-4\,\frac{\beta}{\mathcal{L}} \bigg)+\mathcal{O}(\beta^2).
\end{align}
Figure (\ref{modesincfig}) depicts stationary states $\II$ with $\ell = 0,1,..,5$.

Collecting the above results, the normal inclination modes of a self-gravitating disk are expressed as follows:
\begin{align}
\eta_{\ell}&=c_\ell\,\exp\bigg[-\imath\, \bigg(\frac{\ell\,\pi}{\mathcal{L}} \bigg)^2\bigg(1-4\,\frac{\beta}{\mathcal{L}} \bigg)\,\omega_i\, t \bigg] \nonumber \\
&\times \bigg[ \cos\bigg( \frac{\ell\,\pi\,\rho}{\mathcal{L}}\sqrt{1-4\,\frac{\beta}{\mathcal{L}} } \bigg) - \beta\,\frac{\ell\,\pi}{\mathcal{L}} \nonumber \\
&\times \sqrt{1-4\,\frac{\beta}{\mathcal{L}}}\, \sin \bigg( \frac{\ell\,\pi\,\rho}{\mathcal{L}}\sqrt{1-4\,\frac{\beta}{\mathcal{L}}} \bigg) \bigg].
\end{align}
We note that this solution corresponds to a special case of constant $\omega_i$, which is in turn facilitated by a particular choice of the surface density profile (equation \ref{sigmaprofile}). Employing the $\beta\rightarrow0$ approximation once again, we obtain the simple expression:
\begin{align}
\eta_{\ell}\approx c_\ell\,\exp\bigg[-\imath\, \bigg(\frac{\ell\,\pi}{\mathcal{L}} \bigg)^2\,\omega_i\, t \bigg] \cos\bigg( \frac{\ell\,\pi\,\rho}{\mathcal{L}}\bigg).
\label{eigeninc}
\end{align}

Within the framework of this solution, the coefficients $c_\ell$ are determined from Fourier decomposition of the initial conditions. Upon determination of these quantities, superposition of the eigenstates (\ref{eigeninc}) fully describes the secular evolution of a self-gravitating disk. Physically, these eigenstates describe stationary nodal bending waves, with regression frequencies given by $\omega_\ell$. A number of low-frequency modes are shown in Figure (\ref{incmodesphysfig}), as they appear in physical space. Note that unlike the quantum infinite square well, where the ground state of the wavefunction corresponds to $\ell=1$, the lowest index allowed within the context of our formalism is $\ell=0$, which describes a uniformly inclined, static disk.

\subsubsection{Beyond Nearest Neighbors} \label{BNN}

The preceding analysis was carried out entirely within the limited framework of the nearest-neighbor interactions. Let us now quantify the validity of this assumption. We begin by extending the range of interactions, such that wire $j$ can now interact with neighbors up to index $j\pm2$. The relevant disturbing function of the discrete system is then:
\begin{align}
\RR_j &= B_{jj} \, \eta_j \,\eta^*_j + B_{jj-1} \, \big(\eta_j \,\eta^*_{j-1}+\eta_{j-1} \,\eta^*_{j} \big) \nonumber \\
&+ B_{jj+1} \, \big(\eta_j \,\eta^*_{j+1}+\eta_{j+1} \,\eta^*_{j} \big) + B_{jj-2} \, \big(\eta_j \,\eta^*_{j-2} \nonumber \\ 
&+\eta_{j-2} \,\eta^*_{j} \big) + B_{jj+2} \, \big(\eta_j \,\eta^*_{j+2}+\eta_{j+2} \,\eta^*_{j} \big).
\label{Rieta2}
\end{align}

For the sake of this demonstration, let us assert that because $\beta\ll1$, all of the annuli in question are still in sufficient proximity to one-another for the $\alpha\rightarrow1$ limit to apply. Then, we can crudely assume that most of the variation among the coefficients $B$ will stem from evaluation of the Laplace coefficient at different values of $\alpha$. Thus, setting $\alpha=1/(1+\beta)^\nu$, where $\nu$ is an integer that indexes the interaction length (i.e. $j$:$j\pm\nu$ coupling), we have
\begin{align}
\alpha\,\tilde{b}_{3/2}^{(1)} \approx \frac{2}{(1+\nu^2)\,\pi\,\beta^2} + \mathcal{O}(\beta^{-1}),
\label{feexp2}
\end{align}
which yields $B_{jj\pm2}\approx(2/5)\,B_{jj\pm1}$.

Following up on the same procedure as before, we deduce the equation of motion for wire $j$:
\begin{align}
\frac{d\eta_j}{dt} &\approx\imath\,\bigg( \frac{n_j}{4\,\pi}\frac{m_j}{M}\frac{1}{\beta^2} \bigg)\bigg[ -\bigg(2+\frac{4}{5} \bigg)\eta_j+\eta_{j+1}+\eta_{j-1} \nonumber \\
&+ \frac{2}{5}\bigg(\eta_{j-2} + \eta_{j+2} \bigg) \bigg].
\label{beyond2}
\end{align}
The new terms give rise to a higher-order derivative, such that the continuum limit takes the form:
\begin{align}
\frac{\partial\eta}{\partial t} &\approx \imath\,\omega_i\,\bigg[ \frac{13}{5} \frac{\partial^2\eta}{\partial\rho^2}+\beta^2\,\frac{2}{5}\frac{\partial^4\eta}{\partial\rho^4}  \bigg].
\label{beyond2cont}
\end{align}
Continuing this procedure to $j\pm3$ yields
\begin{align}
\frac{\partial\eta}{\partial t} &\approx \imath\,\omega_i\,\bigg[\frac{22}{5} \frac{\partial^2\eta}{\partial\rho^2}+\beta^2\,\frac{8}{5}\frac{\partial^4\eta}{\partial\rho^4} +\beta^4\,\frac{1}{5}\frac{\partial^6\eta}{\partial\rho^6}  \bigg]
\label{beyond3cont}
\end{align}
and so on.


\begin{figure*}
\centering
\includegraphics[width=0.85\textwidth]{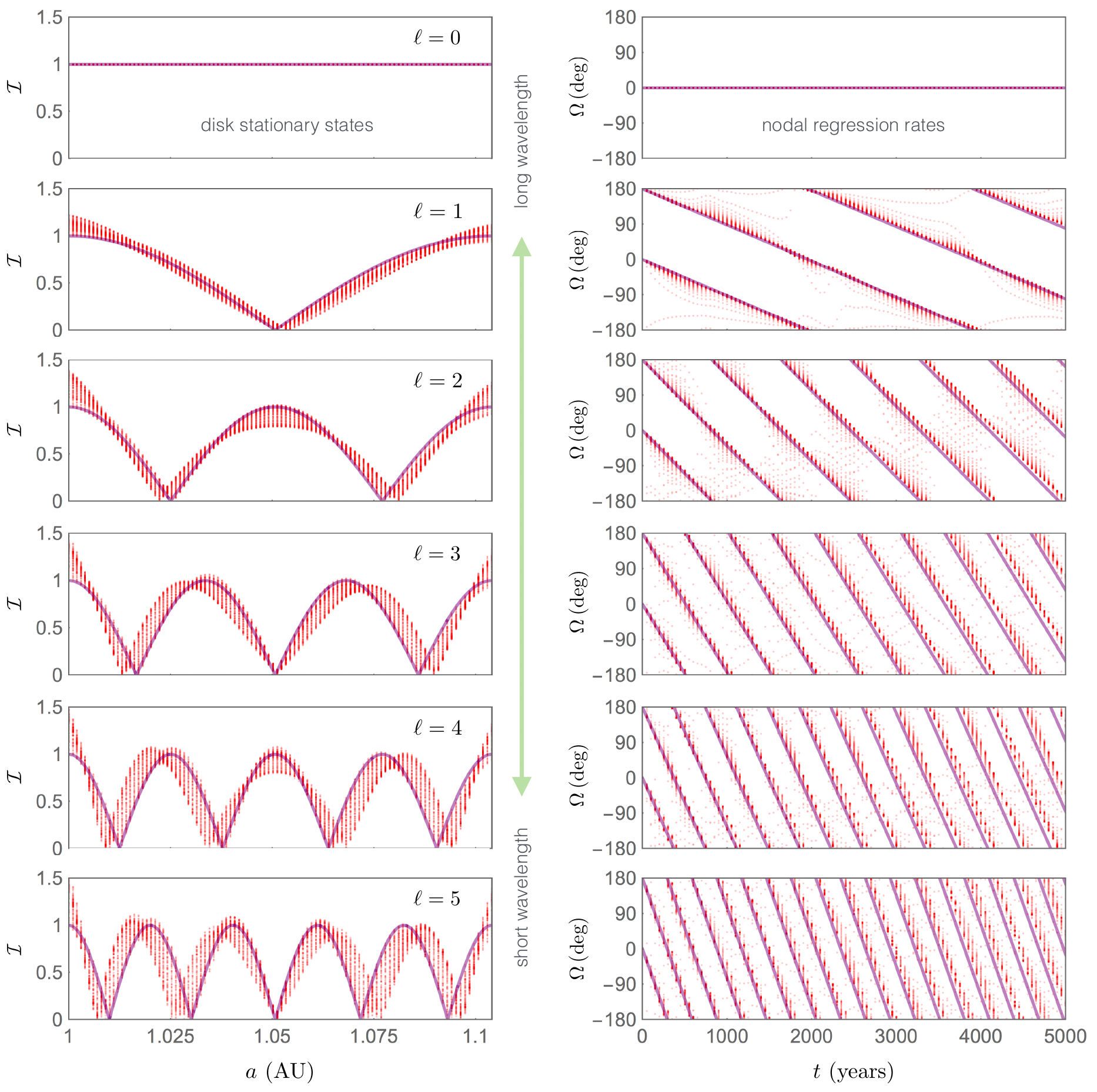}
\caption{Comparison between a full Lagrange-Laplace model of a self-gravitating disk (red dots) and analytical theory based upon the continuum treatment presented herein (purple lines). The left panels depict stationary $\ell=0,1,...,5$ inclination modes of the disk as a function of semi-major axis, and the right panels show the corresponding rates of nodal regression. Over secular timescales, the discrete solution experiences low-amplitude oscillations around its assumed initial conditions, but nevertheless remains approximately confined to the derived eigenstates of the system. In this example, we adopted a gravitational softening parameter of $\beta=10^{-3}$, and considered a disk composed of $N=100$ wires. The disk was taken to orbit a $M=1M_{\odot}$ star, comprise $m_{\rm{disk}} = 1M_{\oplus}$ in total, and extend from $a_{\rm{in}}=1\,$AU to $a_{\rm{out}}=(1+\beta)^N\approx1.1\,$AU. Clearly, the agreement between analytical theory and the discrete model is satisfactory, especially for low-frequency modes.}
\label{LLfulltheory}
\end{figure*}

The general form of the above expression is reminiscent of boundary layer analysis of fluid mechanics (see e.g. section IV of \citealt{1959flme.book.....L}), and elucidates that as long as the characteristic wavelength of interest exceeds the gravitational softening length by a significant margin (which is an implicit assumption of the continuum limit), the contributions due to higher-order derivatives can be safely neglected. In other words, the functional form of the eigenstates derived within the framework of the nearest-neighbor interactions represents a perfectly adequate approximation to the true long-wavelength secular modes of a globally coupled system. Intuitively, the validity this approximation within the context of low wavenumber perturbations can be understood as a consequence of the fact that if $2\pi/k\gg\beta$, the complex field $\eta$ does not change dramatically on scales comparable to the softening length, meaning that its relevant features are encapsulated within its curvature to a good approximation. 

\subsubsection{Collective Effects} \label{CEs}
Although the proceeding analysis demonstrates that eigenmodes deduced from Schr\"{o}dinger's equation (\ref{Schrodinc}) apply even when global effects are taken into account (we will check this assertion more thoroughly below), it also illuminates a key shortcoming of the local treatment of the dynamics. Namely, equations (\ref{beyond2cont}) and (\ref{beyond3cont}) clearly suggest that the eigenfrequencies obtained within the context of the nearest-neighbor approximation severely underestimate the true evolution rates of the disk's normal modes, since each incremental increase in interaction range significantly boosts the effective value of $\omega_i$. Accordingly, let us now alleviate this limitation and compare the resulting formulae with a complete Lagrange-Laplace model of a disk.

The specific goals of the following calculations are two-fold. First, by initializing the full Lagrange-Laplace solution in eigenstates derived from the Schr\"{o}dinger's equation and examining the resulting temporal evolution, we will determine how close these functions are to the true normal modes of the fully coupled system. That is, if the coherence of the initial condition is retained to a good approximation as time marches forward, the initial state is an eigenfunction of the system. On the other hand, if the inclinations oscillate with a large amplitude, then the assumed initial condition cannot be considered a stationary mode of the disk. Second, we will obtain the nodal regression rates corresponding to eigenfunctions (\ref{eigeninc}), fully accounting for long-range gravitational interactions. For succinctness, we will obtain the analytic expression for the evolution rates first.

In order to properly describe collective effects within the system, let us return to a discrete representation of the dynamics of wire $j$, accounting for the potential of the full disk. The appropriate form for the disturbing function reads \citep{MD99}:
\begin{align}
\RR_j &= \frac{1}{2}B_{jj} \, (p_j^2+q_j^2) + \sum_{k=1,j\ne k}^{N} B_{jk} (p_{j}\,p_{k}+q_{j}\,q_{k}).
\label{RRfull}
\end{align}
Recalling the definitions of $p$ and $q$ from equations (\ref{pq}), the expression for the longitude of ascending node is:
\begin{align}
\Omega=\arctan\bigg(\frac{p}{q}\bigg).
\label{Omegapq}
\end{align}

Since we are interested in the the evolution rate of a specific mode, we can readily assume that a common phase (modulo sign) is shared by all constituent annuli within the disk, and its rate of change is independent of its value. Thus, without loss of generality we can set $p=0$ everywhere, which yields a simplified expression for the derivative:
\begin{align}
\frac{d\Omega_j}{dt}\bigg|_{p=0}&=\frac{1}{q_j}\frac{dp_j}{dt}=\frac{1}{q_j}\frac{\partial\RR_j}{\partial q_j} \nonumber \\
&= \frac{1}{q_j}\Bigg(B_{jj}q_j + \sum_{k=1,j\ne k}^{N} B_{jk} q_{k} \Bigg).
\label{dOmegadt}
\end{align}

In similarity with equations (\ref{Bcoeffs}), the interactions coefficients take the form:
\begin{align}
&B_{jj}=-\frac{n_j}{4}\sum_{k=1,j\ne k}^{N}\frac{m_{k}}{M}\,\alpha_{jk}\bar{\alpha}_{jk}\,\tilde{b}_{3/2}^{(1)}\{ \alpha_{jk} \} \nonumber \\
&B_{jk}=\frac{n_j}{4}\frac{m_{k}}{M}\,\alpha_{jk}\bar{\alpha}_{jk}\,\tilde{b}_{3/2}^{(1)} \{ \alpha_{jk} \}
\label{BjjBjk}
\end{align}
where we have adopted the notation of \citet{MD99} to set $\alpha_{jk}=(a_j/a_k)$, $\bar{\alpha}_{jk}=(a_j/a_k)$ if $j<k$ (external perturbation) and $\alpha_{jk}=(a_j/a_k)$, $\bar{\alpha}_{jk}=1$ if $j>k$ (internal perturbation). To this end, note that since we are no longer working within the confines of nearest-neighbor interactions, expression (\ref{feexp}) does not apply as a valid estimate of $\tilde{b}_{3/2}^{(1)}$, meaning that Laplace coefficients must be evaluated at different values of $\alpha_{jk}$ explicitly\footnote{For computational ease, it is useful to express $\tilde{b}_{3/2}^{(1)}$ in terms of standard elliptic integrals \citep{2003ApJ...595..531H}.}.

Substituting the $\beta\rightarrow0$ solution (\ref{eigeninc}) for $q$, i.e.
\begin{align}
q^\ell_j = \cos\bigg( \frac{\ell \pi}{\mathcal{L}} \log \bigg(\frac{a_j}{a_0} \bigg) \bigg),
\label{ql}
\end{align}
and taking the continuous limit, the scaled precession rate at a given semi-major axis $\ahat$ is given by
\begin{align}
\bigg(\frac{d\hat{\Omega}}{dt} \bigg)_\ell&= -\frac{\hat{n}}{4} \bigg( \int_{a_{\rm{in}}}^{\ahat(1-\beta/2)} \frac{\bar{\Sigma} \abar}{M} \, \bigg( \frac{\abar}{\ahat} \bigg) \, \tilde{b}_{3/2}^{(1)} \bigg\{ \frac{\abar}{\ahat} \bigg\} d\abar \nonumber \\
&- \int_{a_{\rm{in}}}^{\ahat(1-\beta/2)} \frac{\bar{\Sigma} \abar}{M} \, \bigg( \frac{\abar}{\ahat} \bigg) \, \tilde{b}_{3/2}^{(1)} \bigg\{ \frac{\abar}{\ahat} \bigg\} \frac{\cos\bigg(\frac{\ell \pi}{\mathcal{L}} \log\bigg(\frac{\abar}{a_0} \bigg)\bigg)}{\cos\bigg(\frac{\ell \pi}{\mathcal{L}} \log\bigg(\frac{\ahat}{a_0} \bigg)\bigg)} d\abar \nonumber \\
&- \int_{\ahat(1+\beta/2)}^{a_{\rm{out}}} \frac{\bar{\Sigma} \abar}{M} \, \bigg( \frac{\ahat}{\abar} \bigg)^2 \, \tilde{b}_{3/2}^{(1)} \bigg\{ \frac{\abar}{\ahat} \bigg\} \frac{\cos\bigg(\frac{\ell \pi}{\mathcal{L}} \log\bigg(\frac{\abar}{a_0} \bigg)\bigg)}{\cos\bigg(\frac{\ell \pi}{\mathcal{L}} \log\bigg(\frac{\ahat}{a_0} \bigg)\bigg)} d\abar \nonumber \\
&+\int_{\ahat(1+\beta/2)}^{a_{\rm{out}}}  \frac{\bar{\Sigma} \abar}{M} \, \bigg( \frac{\ahat}{\abar} \bigg)^2 \, \tilde{b}_{3/2}^{(1)} \bigg\{ \frac{\abar}{\ahat} \bigg\} d\abar \bigg),
\label{dOmegadtellcont}
\end{align}
where the Laplace coefficients (equation \ref{Laplacecoeff}) are evaluated setting $\alpha$ to the semi-major axis ratio inside the curly brackets, and the hatted and barred quantities are evaluated at $\hat{a}$ and $\bar{a}$ respectively. Assuming perfect rigidity of the mode under consideration, we assert that the true evolution rate is given by the average of the regression rates of the constituent annuli of the disk, weighted by their angular momentum deficit:
\begin{align}
\bar{\omega}_\ell &= \int_{a_{\rm{in}}}^{a_{\rm{out}}} \bigg(\frac{d\hat{\Omega}}{dt} \bigg)_\ell \frac{\hat{\Sigma} \ahat}{M} \sqrt{\G M \ahat} \cos^2\bigg(\frac{\ell \pi}{\mathcal{L}} \log\bigg( \frac{\ahat}{a_0} \bigg) \bigg) d\ahat \nonumber \\
&\times \bigg( \int_{a_{\rm{in}}}^{a_{\rm{out}}} \frac{\hat{\Sigma} \ahat}{M} \sqrt{\G M \ahat} \cos^2\bigg(\frac{\ell \pi}{\mathcal{L}} \log\bigg( \frac{\ahat}{a_0} \bigg) \bigg) d\ahat \bigg)^{-1}.
\label{Omegaitrue}
\end{align}

In order to evaluate the validity of the above equations quantitatively, we compare the derived analytic theory to the full Lagrange-Laplace solution of a disk comprised of $N=100$ discrete annuli\footnote{Despite being analytic in nature, the actual solution of a Lagrange-Laplace system with the aid of computer algebra becomes computationally taxing for $N$ substantially greater than $\sim100$.}. Specifically, adopting a softening parameter of $\beta=10^{-3}$ as in \citep{KocsisTremaine2011} and following the recipe outlined in section \ref{sectdiskprofile}, we initialize the disk in a ``pure'' state given by equation (\ref{eigeninc}). Obtaining the solution via the standard approach of matrix diagonalization (see e.g. Ch. 7 of \citealt{morbybook}), we examine the extent to which the disk remains in the prescribed stationary state, and compare the global evolution rate to that dictated by equation (\ref{Omegaitrue}).

Figure (\ref{LLfulltheory}) shows the comparison between the discrete Lagrange-Laplace model of the disk and the analytical theory described above for $\ell=0,1,...,5$. Specifically, the left panel of the Figure depicts the oscillations amplitude of inclination as a function of semi-major axis of the full Lagrange-Laplace system (red dots) over many secular precession timescales. The right panel depicts temporal evolution of the node, wherein values corresponding to all orbital radii are plotted together as a function of time.

Clearly, the eigenstates derived from Schr\"{o}dinger's equation provide an adequate approximation to the global behavior of the simulated system, since the full solution never drifts away from its initial condition (purple curves on the left panel), and instead only experiences low-amplitude oscillations around the derived eigenfunctions. Moreover, the analytically computed evolution rates (purple lines on the right panel) also match the discrete solution well. Accordingly, we conclude that a linear superposition of normal modes of the form:
\begin{empheq}[box=\fbox]{align}
\eta_{\ell}\approx c_\ell\,\exp\bigg[\imath\, \bar{\omega}_\ell \, t \bigg] \cos\bigg( \frac{\ell\,\pi\,\rho}{\mathcal{L}}\bigg).
\label{eigenincbar}
\end{empheq}
provides a simple and easily computable avenue towards evaluating the evolution of geometrically thin, self-gravitating disks on secular timescales.

\section{Perturbed Disks}\label{sect4}

In the last section, we analyzed the dynamics of isolated self-gravitating disks, in the continuum limit of the Lagrange-Laplace secular theory. While examples of such cloistered systems surely exist in nature, real astrophysical disks often reside in phenomenologically rich, dynamic environments, and can be subject to substantial external (or internal) perturbations. Extending the formalism developed above to account for a select class of such perturbations is the primary goal of this section. For definitiveness, here we restrict ourselves to considerations of external gravitational forcing, although it is understood that other extrinsic effects (e.g. radiative stripping, turbulent infall of material, etc) can also influence astrophysical disks' long-term evolution. As in the proceeding section, we concentrate on strictly secular perturbations.

As already discussed above, secular interactions generically arise as a consequence of the orbit-averaged gravitational field of bound companions. Within the context of circumstellar disks, such companions can be binary stars \citep{Batygin2012,2014MNRAS.440.3532L}, or massive planets \citep{2017AJ....153...60M,Nesvold2017}. Under appropriate conditions, the ambient potential of a stellar birth cluster can also be modeled in this manner. In the Galactic center, perturbations due to a molecular torus residing at $\sim1.5-7\,$pc \citep{2005ApJ...622..346C} as well as the gravitational effects of a putative intermediate-mass black hole residing outside the stellar disk \citep{2007ApJ...666..919Y} can be treated within the secular framework. Given the well-known limitations of Lagrange-Laplace theory, we do not aim to provide a complete description of secular dynamics that covers every imaginable regime. Instead, here we focus on deriving a quantitative measure of the disk's tendency towards deformation in face of external excitations.

\subsection{Secular Forcing}

Consider a perturbing companion of mass $m'$, residing on an arbitrarily inclined orbit with eccentricity $e'$ and semi-major axis $a'\gg a_{\rm{out}}$. Envision that the angular momentum of the companion greatly exceeds the angular momentum budget of the disk, such that the back-reaction of the disk upon the companion can be neglected. Under these assumptions, we may orient the coordinate system to coincide with the plane of the companion's orbit (such that $i'=0$) and expand the gravitational potential in powers of the semi-major axis ratio\footnote{Unlike the ``literal'' expansion of the disturbing function employed within the framework of the Lagrange-Laplace theory (which assumes small eccentricities and inclinations while placing no restrictions upon the semi-major axis ratio), expansion of the potential in terms of the semi-major axis ratio assumes that $(a/a')\ll1$ but places no restrictions on the orbital eccentricities and inclinations.}, $(a/a')$ \citep{Kaula1962AJ.....67..300K}. To quadrupole order\footnote{For an extensive exploration of secular dynamics governed by higher order terms, see e.g. \citet{2013MNRAS.431.2155N,2014ApJ...791...86L}.}, the secular disturbing function associated with the orbit-averaged gravitational potential of the companion has the form \citep{2010MNRAS.407.1048M}:
\begin{align}
\mathcal{R}&=\frac{n}{4} \frac{m'}{M}\bigg(\frac{a}{a'}\bigg)^3\frac{1}{\sqrt{1-e'^2}^3} \bigg[\bigg(1+\frac{3}{2}e^2 \bigg)\bigg(\frac{3\,\cos^2(i)-2}{2} \bigg)\nonumber \\
&+\frac{15}{4}e^2\,\sin^2(i)\,\cos(2(\varpi-\Omega)) \bigg].
\label{Rsecquad}
\end{align}

The harmonic term that appears on the second line of equation (\ref{Rsecquad}) governs the Kozai-Lidov effect \citep{1962P&SS....9..719L,1962AJ.....67..591K}, a flavor of dynamical evolution that can manifest as coupled large-scale oscillations of eccentricity and inclination. A number of recent works \citep{2014ApJ...792L..33M,2015ApJ...813..105F,Nesvold2017,2017MNRAS.467.1957Z} have studied the Kozai-Lidov effect pertaining to astrophysical disks, and have shown that it can indeed operate under specific conditions. Here, we assume that these conditions are not satisfied. Practically, this assumption can be translated to a restriction on the mutual disk-companion inclination (namely $i<\arccos(\sqrt{3/5})$). However, we also note that even in disks whose inclination exceeds this critical value, the Kozai effect is typically suppressed due to self-induced regression of the argument of perihelion \citep{Batygin2011,2014MNRAS.440.1179X}. In such instances, this term can be readily averaged away and discarded from the disturbing function. 


Defining the scaled \Poincare\ action-angle variables \citep{morbybook}
\begin{align}
&\tilde{Z}=\sqrt{1-e^2}\big(1-\cos(i) \big) \approx 1-\cos(i) &\tilde{z}=-\Omega,
\label{Poincare}
\end{align}
and assuming that $e\ll1$ as before, the disturbing function (\ref{Rsecquad}) becomes
\begin{align}
\mathcal{R}&=\frac{n}{4}\frac{m'}{M}\bigg(\frac{a}{a'}\bigg)^3\bigg( \frac{1-3\,\tilde{Z}+3\,\tilde{Z}^2/2}{\sqrt{1-e'^2}^3} \bigg).
\label{Rsecquadact}
\end{align}
The resulting nodal regression rate is given by
\begin{align}
&\frac{d\Omega}{dt}=\frac{\partial \mathcal{R}}{\partial \tilde{Z}}\approx-\frac{3}{4}\frac{m'}{M} \bigg(\frac{a}{a'}\bigg)^3\frac{n}{\sqrt{1-e'^2}^3},
\label{omegaprime}
\end{align}
where we have taken the mutual inclination to be small enough for the approximation $\cos(i)\approx1$ to hold.


Equation (\ref{omegaprime}) describes phase rotation, the rate of which is independent of $\eta$. Thus, carrying out the analysis outlined in the previous section, it is immediately evident that this effect can be incorporated into Schr\"{o}dinger's equation as a potential term. Recalling the definition of the logarithmic coordinate, $\rho$ (equation \ref{rho}), we have:
\begin{align}
\imath\, \omega_i\,\frac{\partial \eta}{\partial t} = - \omega_i^2\, \frac{\partial^2\eta}{\partial \rho^2} +\omega_i\,\omega'\,\exp(3\rho/2)\,\eta.
\label{schrodpert}
\end{align}
Maintaining the definitions outlined in section (\ref{sectdiskprofile}), in the above expression, we have scaled the perturbation constant by the parameters relevant to the inner edge of the disk:
\begin{align}
\omega'=\frac{3}{4} \frac{m'}{M} \sqrt{\frac{\G\,M}{a_{\rm{in}}^3}} \bigg( \frac{a_{\rm{in}}}{a'} \bigg)^3 \frac{1}{\sqrt{1-e'^2}^3}.
\label{omegaprimeexp}
\end{align}
In the context of quantum mechanics, a Schr\"{o}dinger's equation with an exponential potential constitutes a rudimentary model for molecular interactions \citep{1982JPhB...15.2689A,2008PhLA..372.3149A}. Therefore, equation (\ref{schrodpert}) implies that gravitational perturbations exerted upon a quasi-Keplerian astrophysical disk can be understood within a framework that is closely related to quantum scattering theory.  

Employing the standard separation of variables (\ref{etasol}) once again, we obtain the following equation for the stationary states:
\begin{align}
-\omega'\exp(3\rho/2)\,\II+\omega_\ell\, \II+\omega_i\,\frac{\partial^2\II}{\partial\rho^2}=0.
\label{sigmafun}
\end{align}
A solution to equation (\ref{sigmafun}) was first identified\footnote{Through a change of variables, equation (\ref{sigmafun}) can be turned into Bessel's equation.} by \citet{Ma1946}, under the condition that $\II$ must vanish at the origin (see also \citealt{1949RvMP...21..488B}). Although a useful starting point, Ma's solution is not directly applicable to secular disk dynamics. Instead, the boundary conditions relevant to the problem at hand are obtained using the procedure outlined in section \ref{BCS}:
\begin{align}
&\frac{\partial \II}{\partial \rho} = -\beta\bigg(\frac{\omega_\ell}{\omega_i} - \frac{\omega'}{\omega_i} \bigg)\II \ \bigg|_{\rho=0} \nonumber \\
&\frac{\partial \II}{\partial \rho} =  \beta\bigg(\frac{\omega_\ell}{\omega_i} - \frac{\omega'}{\omega_i}\exp (3\mathcal{L}/2) \bigg) \II \ \bigg|_{\rho=\mathcal{L}}.
\label{sigmaboundsgr}
\end{align}

Together, equations (\ref{sigmafun}) and (\ref{sigmaboundsgr}) admit an analytic solution, which is expressed in terms of the following quantity 
\begin{align}
\Lambda=&\Gamma\bigg[1+\frac{4\,\imath}{3}\sqrt{\frac{\omega_\ell}{\omega_i}} \bigg] \nonumber \\
&\times\Bigg(\mathcal{X}_{\frac{4\,\imath}{3} \sqrt{\frac{\omega_\ell}{\omega_i}}} \bigg( \frac{4}{3} \sqrt{\frac{\omega'\,\exp(3\rho/2)}{\omega_i}} \bigg) \nonumber \\
&-\Bigg[\mathcal{X}_{-\frac{4\,\imath}{3} \sqrt{\frac{\omega_\ell}{\omega_i}}} \bigg( \frac{4}{3} \sqrt{\frac{\omega'\,\exp(3\rho/2)}{\omega_i}} \bigg) \nonumber \\
&\times\bigg(2\,\beta\,(\omega_\ell-\omega')\, \mathcal{X}_{\frac{4\,\imath}{3} \sqrt{\frac{\omega_\ell}{\omega_i}}} \bigg( \frac{4}{3} \sqrt{\frac{\omega'}{\omega_i}} \bigg) \nonumber \\
&+\sqrt{\omega_i\,\omega'}\, \mathcal{X}_{\frac{4\,\imath}{3} \sqrt{\frac{\omega_\ell}{\omega_i}}-1} \bigg( \frac{4}{3} \sqrt{\frac{\omega'}{\omega_i}} \bigg) \nonumber \\
&+\mathcal{X}_{\frac{4\,\imath}{3} \sqrt{\frac{\omega_\ell}{\omega_i}}+1} \bigg( \frac{4}{3} \sqrt{\frac{\omega'}{\omega_i}} \bigg) \bigg) \Bigg] \nonumber \\
&\times \Bigg[ 2\,\beta\,(\omega_\ell-\omega' )\,\mathcal{X}_{-\frac{4\,\imath}{3} \sqrt{\frac{\omega_\ell}{\omega_i}}} \bigg( \frac{4}{3} \sqrt{\frac{\omega'}{\omega_i}} \bigg) \nonumber \\
&+\sqrt{\omega_i\,\omega'}\,\bigg( \mathcal{X}_{\frac{-4\,\imath}{3} \sqrt{\frac{\omega_\ell}{\omega_i}}-1} \bigg( \frac{4}{3} \sqrt{\frac{\omega'}{\omega_i}} \bigg) \nonumber \\
&+ \mathcal{X}_{1-\frac{4\,\imath}{3} \sqrt{\frac{\omega_\ell}{\omega_i}}} \bigg( \frac{4}{3} \sqrt{\frac{\omega'}{\omega_i}} \bigg) \bigg) \Bigg]^{-1}\Bigg),
\label{crazyshit}
\end{align}
where $\Gamma$ is the Gamma function and $\mathcal{X}$ is the modified Bessel function of the first kind. Suitably, the expression for $\II$ reads:
\begin{align}
\II=\Lambda\,\exp\bigg[-\imath \arctan \bigg( \frac{\Im(\Lambda)}{\Re(\Lambda)} \bigg) \bigg].
\label{flexintwice}
\end{align}
It is noteworthy that even though the expression for $\Lambda$ itself is comprised of Bessel functions with imaginary indexes, they are summed together in such a way as to render $\II$ purely real \citep{Bocher1892}.

\begin{figure}
\centering
\includegraphics[width=\columnwidth]{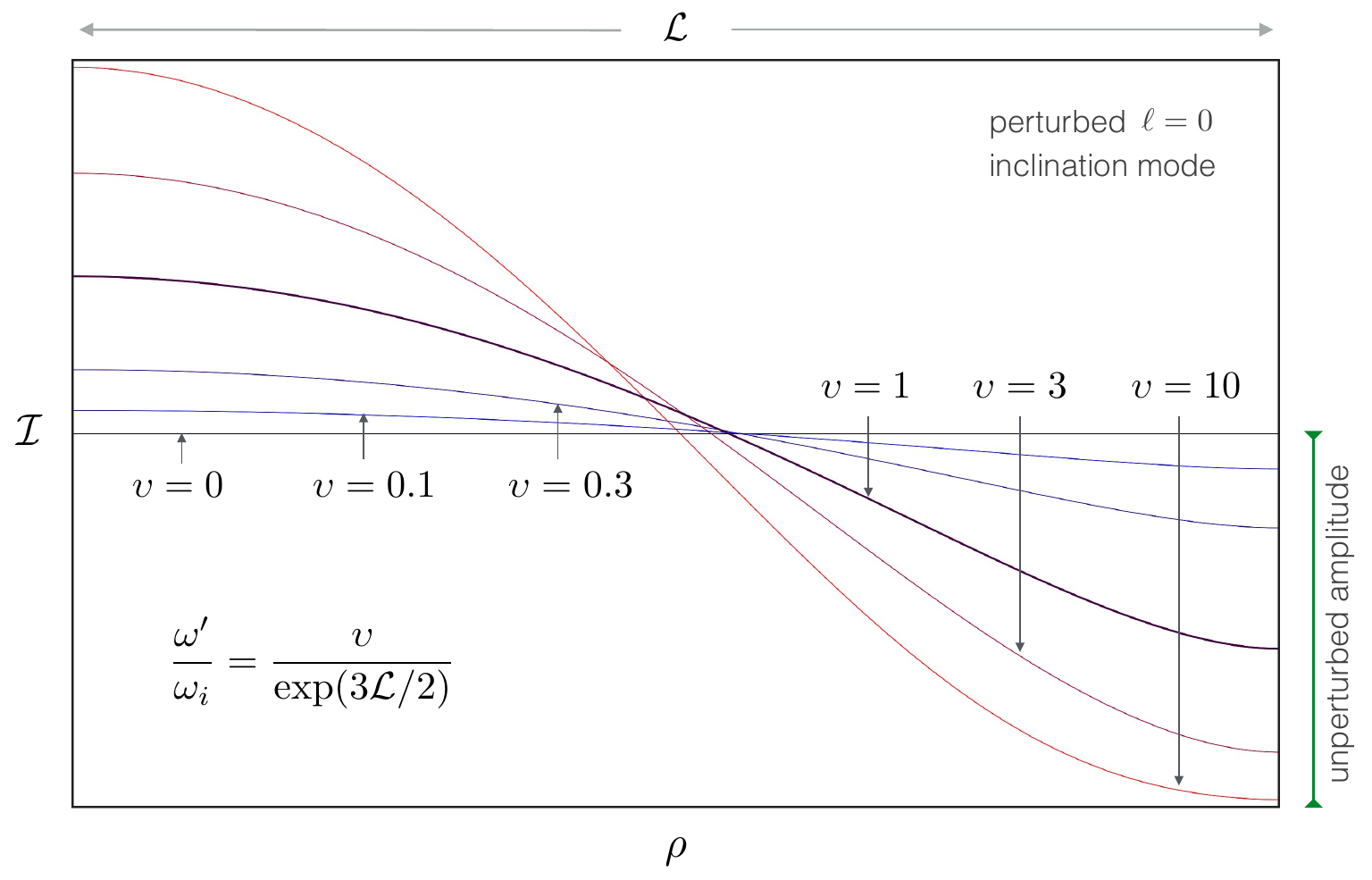}
\caption{Ground state of a perturbed self-gravitating disk, at various secular forcing strengths. As the magnitude of differential precession indued by the external perturber is increased (which corresponds to an increasing $\upsilon$), the ground state of the disk becomes progressively more deformed. Accordingly, a flat, uniformly inclined system is replaced by a globally warped disk. Deformation amplitude of order the unperturbed mode amplitude is attained for $\omega\sim\omega'\exp(3\mathcal{L}/2)$. This condition motivates the criterion for the gravitational rigidity of astrophysical disks presented in section \ref{sectrigid}.}
\label{perturbedfig}
\end{figure}

Unfortunately, the quantization condition that defines the frequencies $\omega_\ell$ for equation (\ref{sigmafun}) is exceedingly cumbersome. Thus, here we restrict ourselves to reporting its approximate form, which is attained by expanding the full expression in $\beta$ to zeroth order:
\begin{align}
&\Bigg[\mathcal{X}_{-1-\frac{4\,\imath}{3}\sqrt{\frac{\omega_\ell}{\omega_i}}}\bigg(\frac{4}{3}\sqrt{\frac{\omega'\,\exp(3\mathcal{L}/2)}{\omega_i}} \bigg)  \nonumber \\
&\times\Bigg(\mathcal{X}_{\frac{4\,\imath}{3}\sqrt{\frac{\omega_\ell}{\omega_i}}-1}\bigg(\frac{4}{3}\sqrt{\frac{\omega'}{\omega_i}} \bigg) +\mathcal{X}_{\frac{4\,\imath}{3}\sqrt{\frac{\omega_\ell}{\omega_i}}+1}\bigg(\frac{4}{3}\sqrt{\frac{\omega'}{\omega_i}} \bigg)\Bigg) \nonumber \\
&+\mathcal{X}_{1-\frac{4\,\imath}{3}\sqrt{\frac{\omega_\ell}{\omega_i}}}\bigg(\frac{4}{3}\sqrt{\frac{\omega'\,\exp(3\mathcal{L}/2)}{\omega_i}} \bigg)  \nonumber \\
&\times\Bigg(\mathcal{X}_{\frac{4\,\imath}{3}\sqrt{\frac{\omega_\ell}{\omega_i}}+1}\bigg(\frac{4}{3}\sqrt{\frac{\omega'}{\omega_i}} \bigg) +\mathcal{X}_{\frac{4\,\imath}{3}\sqrt{\frac{\omega_\ell}{\omega_i}}-1}\bigg(\frac{4}{3}\sqrt{\frac{\omega'}{\omega_i}} \bigg)\Bigg) \nonumber \\
&-\Bigg( \mathcal{X}_{\frac{4\,\imath}{3}\sqrt{\frac{\omega_\ell}{\omega_i}}-1}\bigg(\frac{4}{3}\sqrt{\frac{\omega'\,\exp(3\mathcal{L}/2)}{\omega_i}} \bigg)  \nonumber \\
&+\mathcal{X}_{\frac{4\,\imath}{3}\sqrt{\frac{\omega_\ell}{\omega_i}}+1}\bigg(\frac{4}{3}\sqrt{\frac{\omega'\,\exp(3\mathcal{L}/2)}{\omega_i}} \bigg) \Bigg) \nonumber \\
&\times\Bigg(\mathcal{X}_{-\frac{4\,\imath}{3}\sqrt{\frac{\omega_\ell}{\omega_i}}-1}\bigg(\frac{4}{3}\sqrt{\frac{\omega'}{\omega_i}} \bigg)+\mathcal{X}_{1-\frac{4\,\imath}{3}\sqrt{\frac{\omega_\ell}{\omega_i}}}\bigg(\frac{4}{3}\sqrt{\frac{\omega'}{\omega_i}} \bigg)\Bigg)\Bigg]\nonumber \\
&\times \Gamma\bigg[1-\frac{4\,\imath}{3}\sqrt{\frac{\omega_\ell}{\omega_i}} \bigg] \, \Gamma\bigg[1+\frac{4\,\imath}{3}\sqrt{\frac{\omega_\ell}{\omega_i}} \bigg] =0.
\label{morecrazyshit}
\end{align}
We note that this condition is exact in the limit of a razor-thin disk, and for $(\omega'/\omega_i)$ of order unity or less, the zeroes of this function are well approximated by $\omega_\ell/\omega_i\simeq(\ell\,\pi/\mathcal{L})^2$.

Recall from section \ref{sect3} that the ground state of an unperturbed disk has $\ell=0$, which corresponds to $\eta=\rm{const.}$ everywhere. In presence of external forcing, this simple solution no longer holds, as the nodal regression induced by the companion can lead to a prominent distortion of the lowest energy state. Importantly, a controlling parameter that determines the extent of this distortion is $\upsilon=(\omega'/\omega_i)\exp(3\mathcal{L}/2)$. Physically, this ratio corresponds to the relative magnitudes of the differential precession induced by the companion and nodal regression facilitated by self-gravity.

\begin{figure*}
\centering
\includegraphics[width=0.85\textwidth]{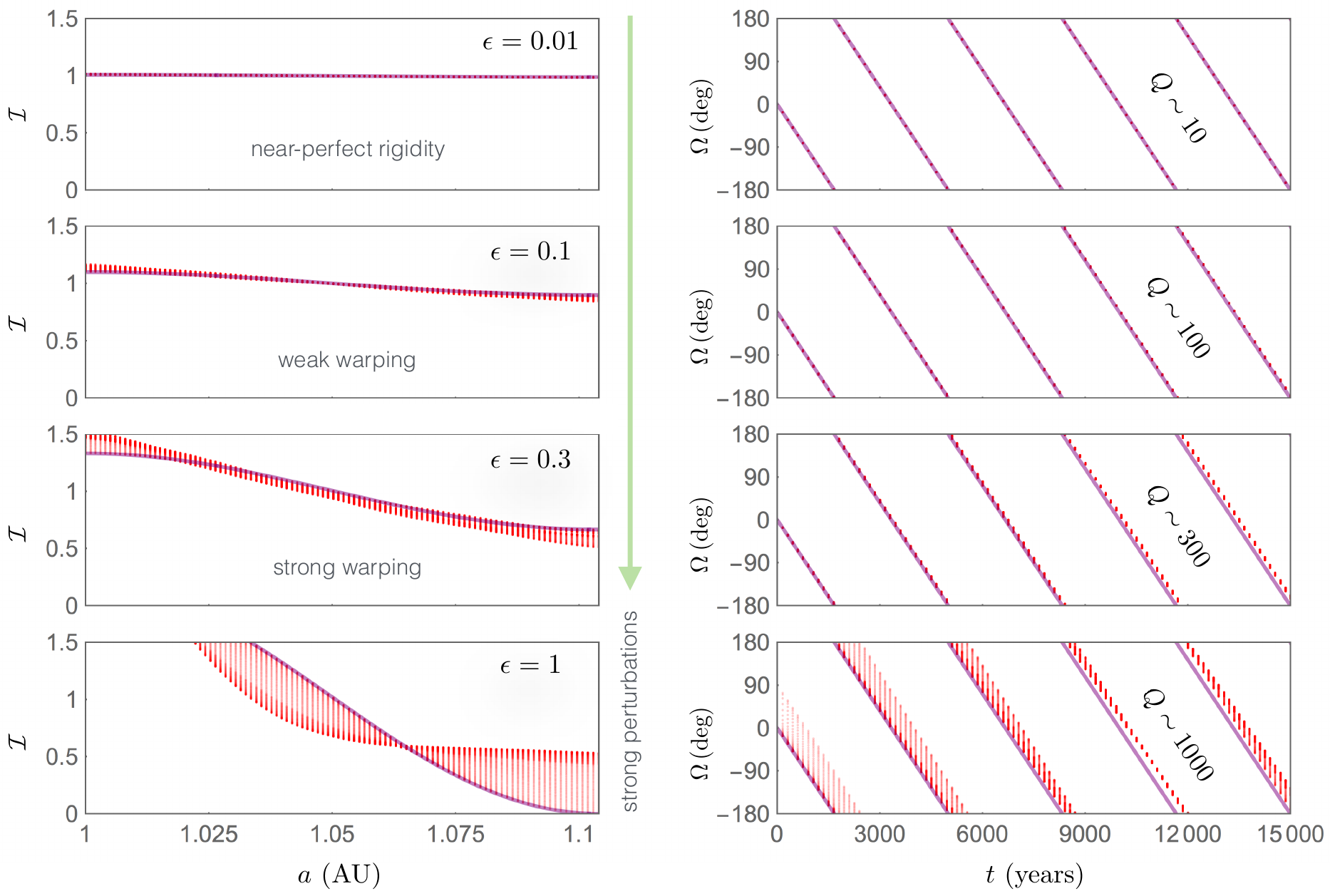}
\caption{Ground states of an externally perturbed self-gravitating disk, corresponding to different values of the deformation parameter, $\epsilon$. As in Figure (\ref{LLfulltheory}), we consider the dynamics of a narrow disk with $\beta=10^{-3}$, composed of $N=100$ massive wires. In the depicted calculations, however, this disk is perturbed by a $m=10^{-2}M$ companion residing at $a'=3\,$AU on a circular orbit. In each panel, the disk is initialized in a state given by equation (\ref{superpos}), and its mass tuned via equation (\ref{ultimatestunt}) to give the desired value of $\epsilon$. As expected from theoretical arguments, the gravitational rigidity of the system is ensured for $\epsilon\lesssim0.1$. When this condition is satisfied, the disk's longitude of ascending node executes a slow regression with respect to the plane defined by the companion's orbit. For reference, the characteristic value of the Toomre's $Q$ within the disk is quoted on the right panel for each realization.}
\label{LLrigid}
\end{figure*}

Forgoing collective effects for the moment, Figure (\ref{perturbedfig}) depicts the lowest frequency mode (as dictated by equation \ref{morecrazyshit}) for various proportions of the aforementioned parameter $\upsilon$. As may be intuitively expected, the lowest-energy eigenfunctions become progressively deformed as the external forcing strength is increased. Moreover, it is worth noting that the extent of deviation away from the unperturbed solution becomes comparable to the amplitude of the unperturbed solution when $\upsilon\sim1$. We will revisit this notion again below. 

\subsection{Gravitational Rigidity}\label{sectrigid}

A key question that can be addressed using the framework derived in the previous subsection is: ``under what conditions, will a disk develop significant structure due to external perturbations?'' However, to answer this question quantitatively, we must re-examine the extent to which the ground state of the perturbed system is deformed by external forcing, accounting for collective effects within the disk.

Figure (\ref{perturbedfig}) shows that to a fair approximation, the functional form of the ground-state of a secularly forced disk can be represented as a super-position of $\ell=0$ and $\ell=1$ states of the unperturbed system:
\begin{align}
\mathcal{I}_0\approx c_0 \bigg[1+\epsilon\,\cos\bigg( \frac{2\,\pi\,\rho}{\mathcal{L}} \bigg) \bigg].
\label{superpos}
\end{align}
In this approximation, the deviation away from strict coplanarity, $\epsilon$, fully encapsulates the strength of external perturbation. The key is thus to compute its magnitude directly from the parameters of the system. 

Following the discussion outlined in section \ref{CEs}, we recall that the evolution rate of the unperturbed $\ell=1$ mode, which we denote as $\bar{\omega}_1$, is given by equations (\ref{dOmegadtellcont}-\ref{Omegaitrue}). Qualitatively, this frequency can be interpreted as the rate of angular momentum redistribution within the disk, facilitated by self-gravity. The resulting value should be compared with the rate of differential nodal regression induced upon the system by the external companion:
\begin{align}
\bar{\omega}'=\omega'\bigg(\bigg(\frac{a_{\rm{out}}}{a_{\rm{in}}}\bigg)^{3/2}-1 \bigg).
\label{superpos}
\end{align}

Guided by the findings of the previous subsection, we argue that the gravitational rigidity of the system is characterized entirely by the relative importance of these two effects, and assert that the dimensionless parameter $\epsilon$ is given by their ratio. Explicitly, in terms of physical parameters of the system, the expression reads:
\begin{align}
\epsilon&= \frac{\bar{\omega}'}{\bar{\omega}_1} = \Bigg[ \int_{a_{\rm{in}}}^{a_{\rm{out}}}  \frac{\G\,M}{4\,\hat{a}}  \frac{\Sigma_0 \,\sqrt{a_0 \, \ahat}}{M} \cos^2\bigg(\frac{\pi}{\mathcal{L}} \log\bigg( \frac{\ahat}{a_0} \bigg) \bigg) \nonumber \\
&\times \Bigg( \int_{a_{\rm{in}}}^{\ahat(1-\beta/2)} \frac{\Sigma_0 \,\sqrt{a_0 \, \abar}}{M} \, \bigg( \frac{\abar}{\ahat} \bigg) \, \tilde{b}_{3/2}^{(1)} \bigg\{ \frac{\abar}{\ahat} \bigg\} d\abar \nonumber \\
&- \int_{a_{\rm{in}}}^{\ahat(1-\beta/2)} \frac{\Sigma_0 \,\sqrt{a_0 \, \abar} \abar}{M} \, \bigg( \frac{\abar}{\ahat} \bigg) \, \tilde{b}_{3/2}^{(1)} \bigg\{ \frac{\abar}{\ahat} \bigg\} \frac{\cos\bigg(\frac{\pi}{\mathcal{L}} \log\bigg(\frac{\abar}{a_0} \bigg)\bigg)}{\cos\bigg(\frac{ \pi}{\mathcal{L}} \log\bigg(\frac{\ahat}{a_0} \bigg)\bigg)} d\abar \nonumber \\
&- \int_{\ahat(1+\beta/2)}^{a_{\rm{out}}} \frac{\Sigma_0 \,\sqrt{a_0 \, \abar}}{M} \, \bigg( \frac{\ahat}{\abar} \bigg)^2 \, \tilde{b}_{3/2}^{(1)} \bigg\{ \frac{\ahat}{\abar} \bigg\} \frac{\cos\bigg(\frac{ \pi}{\mathcal{L}} \log\bigg(\frac{\abar}{a_0} \bigg)\bigg)}{\cos\bigg(\frac{ \pi}{\mathcal{L}} \log\bigg(\frac{\ahat}{a_0} \bigg)\Bigg)} d\abar \nonumber \\
&+\int_{\ahat(1+\beta/2)}^{a_{\rm{out}}}  \frac{\Sigma_0 \,\sqrt{a_0 \, \abar}}{M} \, \bigg( \frac{\ahat}{\abar} \bigg)^2 \, \tilde{b}_{3/2}^{(1)} \bigg\{ \frac{\ahat}{\abar} \bigg\} d\abar \Bigg)\,d\ahat \Bigg] \nonumber \\
&\times \Bigg[\frac{3}{4} \frac{m'}{M} \sqrt{\frac{\G\,M}{a_{\rm{in}}^3}} \bigg( \frac{a_{\rm{in}}}{a'} \bigg)^3 \frac{1}{\sqrt{1-e'^2}^3} \bigg(\bigg(\frac{a_{\rm{out}}}{a_{\rm{in}}}\bigg)^{3/2}-1 \bigg) \nonumber \\
&\times \int_{a_{\rm{in}}}^{a_{\rm{out}}} \frac{\Sigma_0 \,\sqrt{a_0 \, \ahat}}{M} \sqrt{\G M \ahat} \cos^2\bigg(\frac{\pi}{\mathcal{L}} \log\bigg( \frac{\ahat}{a_0} \bigg) \bigg) d\ahat \Bigg]^{-1} \ll 1,
\label{ultimatestunt}
\end{align}
We remind the reader that encoded in this equation (as well as in the form of the solution \ref{crazyshit}-\ref{flexintwice} itself) is the assumption of a $\Sigma\propto1/\sqrt{a}$ surface density profile, meaning that for a different radial dependence of $\Sigma$, the rigidity criterion would take on a quantitatively distinct form.

In order to check that equations (\ref{superpos}) and (\ref{ultimatestunt}) truly represent an adequate approximation to the lowest energy state of a secularly perturbed system, we once again employ the discrete Lagrange-Laplace model of the disk shown in Figure (\ref{LLfulltheory}). Subjecting the disk to secular perturbations arising from a $m'=10^{-2}\,M$ perturber residing on a circular orbit at $(a_{\rm{in}}/a')=1/3$, we tune the disk mass to correspond to $\epsilon=0.01,0.1,0.3$ and $1$, contrasting our analytic results with the full model at each iteration. A variant of Figure (\ref{LLfulltheory}) depicting this comparison is shown in Figure (\ref{LLrigid}). As is made evident by the essentially flat nature of the inclination ground state depicted on the left panels of Figure (\ref{LLrigid}), and the coherence of ascending nodes shown on the right panels, disks characterized by $\epsilon\ll 1$ maintain near-perfect gravitational rigidity, and show only minimal deformation in face of external forcing. 

Qualitatively speaking, the capacity of a disk to retain its unperturbed form when $\epsilon \ll 1$ stems from nothing other than adiabatic invariance of the system's actions \citep{Henrard1993}. Suitably, $\epsilon$ itself represents a readily computable measure of the extent to which the emergent quasi-integrals are conserved. Accordingly, the propensity of self-gravitating near-Keplerian disks to deform under extrinsic forcing is fully encapsulated within this parameter. Just as Toomre's $Q\gtrsim1$ criterion corresponds to a disk that is stable against self-gravitational fragmentation, a perturbed disk with $\epsilon\ll1$ is stable against external secular excitation. In other words, systems with $\epsilon$ substantially smaller than unity will behave as gravitationally rigid bodies.  

\section{Discussion} \label{sect5}

Although the time-dependent Schr\"{o}dinger equation is often thought of as a mathematical description that is reserved for quantum mechanics alone, its nonlinear counterpart (equation \ref{H2}) appears in numerous instances of classical physics. Examples of such contexts include small-amplitude (fluid) gravity waves, Langmuir plasma waves, and nonlinear optics \citep{Bellanbook,Agrawalbook}. In this work, we have shown that the linear Schr\"{o}dinger equation is also keenly relevant to the long-term evolution of astrophysical disks and constitutes a continuum description of the secular dynamics of stable self-gravitating systems. Remarkably, this result stems from the well-known Lagrange-Laplace perturbation theory of celestial mechanics, the origins of which were established over two centuries ago. 

As a perturbative model, the Schr\"{o}dinger equation lends easy access to orbital evolution that unfolds on timescales that greatly exceed the longest Keplerian period of the system, but are nevertheless shorter than its physical lifetime. While the resulting characterization of long-term angular momentum transfer is not readily attainable by other (non-secular) methods, it is of key importance to a judicious interpretation of the observed structure of astrophysical disks. Simultaneously, we emphasize that because of the orbit-averaged framework within which the theory was constructed, it is not designed to capture the full richness of dynamical phenomena that may ensue within self-gravitating quasi-Keplerian systems\footnote{For example, spiral density waves that cause the central object to be displaced from the center of mass and rely upon the resulting indirect potential to further amplify their magnitude \citep{Adams1989} are not captured in our model.}.

By virtue of being exactly solvable, Schr\"{o}dinger's equation provides an enthralling reduction of the full solution of the gravitational $N$-body problem. Within the context of this description, the eigenstates derived from Schr\"{o}dinger's equation translate to normal modes of quasi-Keplerian disks, and the physical interpretation of radially traveling inclination pulses corresponds to propagation of nodal bending waves \citep{BinneyTremaine1987}. The derived formalism further allows us to  formulate a measure of the gravitational rigidity of near-Keplerian systems, subject to extrinsic secular forcing. The resulting $\epsilon \ll1$ stability criterion (equation \ref{ultimatestunt}) quantifies the response of self-gravitating disks to external perturbations, and compliments Toomre's $Q\gtrsim1$ stability criterion for self-gravitational fragmentation. 

In light of our model's simplicity, it is important to keep in mind that Schr\"{o}dinger's equation only provides an adequate description for the angular momentum exchange within self-gravitating disks in a specific parameter regime (i.e., sufficiently low eccentricities and inclinations, small aspect ratio, etc.). That is to say that the reduction of $6N$ nonlinear ordinary differential equations to a single linear partial differential equation necessarily entails some approximations. As a result, the outlined theory cannot serve as a general replacement for more complex numerical simulations \citep{ToumaSoft2009}. Instead, our analytical model can be meaningfully used to provide qualitative context for numerical results. 


There exist numerous ways in which our model can be extended. As a first step, it is straightforward to generalize the derived formalism to account for arbitrary radial dependence of the surface density profile, and to lift the assumption of constant aspect ratio within the disk. While of considerable practical use, these generalizations will endow the fundamental frequency $\omega_i$ with a dependence upon the semi-major axes, compromising the applicability of simple solutions (such as that outlined in equation \ref{eigeninc}) to the governing Schr\"{o}dinger-like equation. 

Another curious extension of our model lies in incorporation of nonlinearity. As already mentioned in section \ref{sect2}, adding non-linear action terms to the governing Hamiltonian yields the non-linear variant of Schr\"{o}dinger's equation in the continuum limit. Importantly, such terms arise naturally in the secular disturbing function at fourth order in eccentricities and inclination \citep{MD99}, and can yield significant coupling among the two degrees of freedom (see e.g. \citealt{2015ApJ...799..120B}). Simultaneously, it is worth noting that adding non-linear terms to the governing Hamiltonian turns the description of self-gravitating disks into a chain of anharmonically coupled oscillators - a system closely related to the Fermi-Pasta-Ulam lattice (see \citealt{1992PhR...213..271F} for a review). As a consequence, it is reasonable to speculate that the resulting evolution will be satisfied by breather solutions, and that Fermi-Pasta-Ulam recurrence can occur in self-gravitating disks. 

Because we have considered purely gravitational coupling in this work, the obtained formulae are only applicable to particle disks, formally speaking. Although the strictly gravitational picture can in some cases serve as a good approximation of a hydrodynamic disk\footnote{Generically, it is reasonable to neglect internal pressure forces if $\sqrt{\G\,\Sigma\,a} \gg c_{\rm{s}}$, where $c_{\rm{s}}$ is the speed of sound \citep{Trem2001}.} (see e.g. \citealt{2010A&A...511A..77F,Batygin2012}), a comprehensive theory for fluid disk evolution must include an account for internal forces. To this end, \citet{2001MNRAS.325..231O,2006MNRAS.365..977O} has demonstrated that the dynamics of a fluid disk subject exclusively to pressure forces (i.e. with negligible viscosity and self-gravity) can be understood within the framework of a nonlinear variant of Schr\"{o}dinger's equation (see also \citealt{2016MNRAS.458.3739B}). Correspondingly, fusing such formalism with the results developed herein may provide a viable way to extend our theory to simultaneously account for gravitational and hydrodynamic effects. 

We conclude this work with a brief discussion of perturbations exerted upon astrophysical disks. As discussed in section \ref{sect4}, the flavor of interactions considered in this manuscript represents only a subset of the full range of possibilities. For example, carrying out the expansion of the secular disturbing function (equation \ref{Rsecquad}) to higher order \citep{2013MNRAS.431.2155N} in the semi-major axis ratio can introduce purely real forcing terms that are independent of $\eta$ nto the governing equations. Along similar lines of reasoning, inclination damping can be modeled as a purely real diffusion term within this framework. On the contrary, stochastic effects, such as those arising from passing stars or turbulent infall of disk material can be simulated by adding appropriately modulated noise to the exterior boundary condition \citep{2014ApJ...797L..29S,2015MNRAS.448..344L}. Taken together, incorporation of aforementioned effects constitutes a intriguing avenue towards further development of the model and a more complete characterization of the long-term evolution of self-gravitating disks. 

\section*{Acknowledgements}
I am grateful to G. Laughlin, F. Adams, D. Stevenson, A. Morbidelli, C. Spalding, S. Michalakis, A. Kitaev, S. Tremaine and J. Touma for illuminating discussions. Additionally, I would like to thank the anonymous referee for providing a thorough and insightful report that has led to a considerable improvement of the manuscript, as well as the David and Lucile Packard Foundation for their generous support.



\bsp	
\label{lastpage}
\end{document}